\documentclass[12pt,preprint]{aastex}
\usepackage{amsmath}
\shorttitle{Density and Rotation in Be~star Disks}
\shortauthors{McGill et al.}
\usepackage{float}
\newcommand{\msun}{\mbox{$M_{\odot}$}}
\newcommand{\rsun}{\mbox{$R_{\odot}$}}
\newcommand{\lsun}{\mbox{$L_{\odot}$}}

\begin{document} 

\title{The Effect of Density on the Thermal Structure of Gravitationally-Darkened Be~Star Disks}

\author{M. A. M\textsuperscript{c}Gill\altaffilmark{1}, T. A. A. Sigut\altaffilmark{1}, \& C. E. Jones\altaffilmark{1}, }

\altaffiltext{1}{Department of Physics and Astronomy, The University 
of Western Ontario, London, Ontario, N6A 3K7, Canada}

\begin{abstract}
The effects of gravitational darkening on the thermal structure of Be~star disks of differing densities are systematically examined.  
Gravitational darkening is the decrease of the effective temperature near the equator and the corresponding increase near the poles of a star caused by rapid rotation.  
We also include the rotational distortion of the star using the Roche Model.  
Increasing the disk density increases the optical depths in the equatorial plane, 
resulting in the formation of an inner cool region near the equatorial plane of the disk.
High rotation rates result in disks that have temperatures similar to those of a denser disk, namely cooler overall.  
However the effect of increasing rotation produces additional heating in the upper disk due to the hotter stellar pole.  
Cool regions in the equatorial plane normally associated with high density are seen in low density models at high rotation rates.  
Gravitational darkening increases the amount of very cool and very hot material in the disk and decreases the amount of disk material at moderate temperatures.  
We also present models which study the effect of  gravitational darkening on hydrostatically-converged disks, 
in which the temperature structure is consistent with vertical hydrostatic equilibrium.  
Because the equatorial regions become cooler, hydrostatically converged models that include gravity darkening 
have smaller vertical scale heights, and $H/R$ is smaller by as much as $56$ \% near $v_{\rm crit}$.  
Finally we explore differences in disk temperatures when alternate formulations 
of gravitational darkening, which lower the temperature difference between the pole and the equator, are used.  
\end{abstract}


\section{Introduction}
\label{intro}

\citet{Jaschek} gives the current working definition of a Be~star as 
``a non‐supergiant B star whose spectrum has, or had at some time, one or more Balmer lines 
in emission.''
The source of these emission lines is a geometrically thin, gaseous disk surrounding the star.  
These disks are often temporary, building up and dissipating over approximately a decade, 
although some Be~stars have shown persistent Balmer emission over their whole observational history \citep{Wisniewski, Carciofi}.  
When present, these disks produce the other characteristic features of a Be~star system.  
(1) A continuum excess occurs due to the cooler disk material \citep{Dougherty}.  
It begins as a weak, but noticeable, excess in the optical, 
resulting in contributions to the (B-V) colour index of about a tenth of a magnitude \citep{Howells}, 
peaking at $\approx 10 \mu m$ and continuing well past the infra-red \citep{wat87}.  
The longest wavelength detection of a Be~star is 6 cm \citep{Dougherty}.  
(2) Continuum polarization is present due to scattering in the non-spherical circumstellar material \citep{Waters1992}.  
(3) Emission occurs in other hydrogen line series, Lyman, Paschen, Brackett, Pfund, and Humphreys \citep{Hony2000,2004Houck_spitzer}.  
(4) Some Be~stars show emission lines in elements other than hydrogen, such as He\,{\sc i}, Fe\,{\sc ii}, and sometimes Si\,{\sc ii} and Mg\,{\sc ii} \citep{Slettebak_lines, porter}.  

The formation mechanism(s) of these disks is still unclear, 
but it is likely facilitated by the rapid rotation 
of these stars which could allow a relatively weak process to drive mass-loss in a 
main sequence or just post-main sequence star \citep{porter}.  
Short term variability, on timescales between 0.5 to 2 days, is observed in the 
photospheric line shapes and photometry of most classical Be stars \citep{Percy}, 
and Be~stars have also been known to exhibit non-radial pulsations, 
indicating disturbances on and above the surfaces of these stars \citep{porter}.  
Disk building appears to a be stochastic process 
occurring in a series of outbursts associated with increased emission \citep{Stefl}.  

The B type stars that produce these disks are rapid rotators \citep{Yudin2001}.  
Their rotation rates are still a matter of debate \citep{cra05, vanBelle2012}, 
but they are fast enough that the effects of rotational distortion 
and gravitational darkening should be considered.  
As described in \S~2 of our previous paper, \citet{me1}, 
gravitational darkening is the rotationally induced reduction of the effective gravity 
towards the stellar equator causing a corresponding decrease in the local temperature.  
The traditional formulation is found in \citet{von_Zeipel_1} and \citet{Collins1963}, 
while newer results are discussed in \citet{vanBelle2012}.  
Due to the small effect of rotation on the total luminosity, 
the pole of a rotating star is actually hotter than an equivalent, non-rotating star.  
Rapid rotation causes the star to become distorted with 
the radius becoming larger at the equator than the pole, to a maximum of 1.5 times the polar radius.  

Interferometric observations of these stars have confirmed a variation in brightness across the surface of rapidly rotating stars \citep{vanBelle2012}.  
The polar regions of such stars are noticeably brighter than the equator \citep{vanBelle2012}.  
However the temperature contrast between the pole and equator is not as strong as expected for the classical formulation of gravitational darkening \citep{vanBelle2012}.  
This is parameterized by $\beta$, as $T_{\rm{eff}} \propto g_{\rm{eff}}^{\beta}$.  
The canonical value of $\beta$ is $0.25$ \citep{von_Zeipel_1}.\footnote{\citet{von_Zeipel_1}  defines $\beta=0.25$ for radiative energy transport.  Convective transport occurs at $ T_{\rm eff} <  \, \sim \! 7000 \, {\rm K}$ which is less than 5 \% of the total area our B5V model rotating at $0.99 \, v_{\rm crit}$ so neglecting this should have no effect on our results.}  
Smaller values of $\beta$ indicate a smaller temperature difference between the pole and the equator.  
\citet{Che2011} find their interferometric observations of Regulus can be best explained by $\beta=0.188$.  
Observationally determined values for $\beta$ are typically between 0.25 and 0.18, and \citet{vanBelle2012} gives $\beta=0.21$ as the typical value seen from interferometry.  
Interferometric data seems to suggest that this only affects the temperature profile, while the shape of the star is consistent with a Roche Model \citep{vanBelle2012}.  

As the star is the energy source for the disk, changes introduced by rotation have an effect on the disk.  
As described in \citet{me1}, the effects of gravitational darkening on models of classical Be~stars is to 
reduce the temperature in the mid-plane of the disk while causing some temperature increases in the upper disk.  
\citet{me1} presented models for four different spectral types: B0V, B2V, B3V, and B5V at ten different rotation rates.  
However only a single disk density scale of $\rho_o=5.0\, \times \, 10^{-11}$ g cm$^{-3}$ (see Equation~(\ref{rho_disk})~) was considered.  
In this paper, we examine the combined effects of variations in disk density and stellar rotation.  
We also include discussions on the effects of rotation on hydrostatically converged models and on the inclusion of different formulations of gravitational darkening.  
The organization of the paper is as follows: 
\S~\ref{calc} briefly outlines the models presented in this paper; 
\S~\ref{results} provides our results; 
\S~\ref{diff_density} explores the effects of rotation when combined with changes in the density of the disk; 
\S~\ref{hydro} looks at the effects of rotation on disk models that have been hydrostatically converged; 
\S~\ref{diff_grav_dark} examines the effects of using different formulations of gravity darkening on the disk temperatures; 
and conclusions are presented in \S~\ref{conclusions}.  

\section{Calculations}
\label{calc}

\begin{table}[h]
\caption{Adopted Stellar parameters \label{star_para}}
\vspace{-0.08in}
\begin{center}
\begin{tabular}{c c c c c c c }
\hline \hline
Type&Mass&Polar Radius&Luminosity&  $T_{\rm{eff}}$  &${\omega}_{\rm{crit}}$ & $ v_{\rm{crit}}$ \\
&(\msun)&(\rsun)& (\lsun) & (K) &(${\rm{rad}}$ ${\rm{s}}^{-1}$)  &  (${\rm{km}}$ ${\rm{s}}^{-1}$) \\ \hline
B0V&17.5&7.40&$3.98\, {\rm x}\, 10^{4}$&30000&$ 7.10 \, {\rm x} \, 10^{-5}$ & $ 548$ \\
B2V&9.11&5.33&$4.76\, {\rm x}\, 10^{3}$&20800&$ 8.38 \, {\rm x} \, 10^{-5}$ & $ 466$ \\
B3V&7.60&4.80&$2.58\, {\rm x}\, 10^{3}$&18800&$ 8.95 \, {\rm x} \, 10^{-5}$ & $ 449$ \\
B5V&5.90&3.90&$7.28\, {\rm x}\, 10^{2}$&15200&$ 1.08 \, {\rm x} \, 10^{-4}$ & $ 439$ \\ \hline
\end{tabular}
\end{center}
\vspace{-0.1in} 
{\it Notes:} ${\omega}_{\rm{crit}}=\sqrt{{ 8 G M }/{27 r_{\rm {p}}^{\,3}}}$, where $r_{\rm{p}}$ is the polar radius of the star and $M$ is its mass.  \\
$v_{\rm{crit}}= r_{\rm{max}} \, \omega_{\rm{crit}}$ , where $r_{\rm{max}}=1.5 \, r_{\rm{p}}$. 
Stellar parameters adopted from \citet{cox2000}.  
\vspace{-0.04in} 
\end{table}

The modelling program, {\sc Bedisk}, was used \citep{sig07}.  {\sc Bedisk} solves the statistical equilibrium equations for the atomic level populations 
and then enforces radiative equilibrium at each point of the computational grid representing the disk.  
The version of {\sc Bedisk} described in \citet{me1} includes gravitational darkening and was run for the stellar 
parameters given in Table~\ref{star_para} and the rotation rates given in Table~\ref{v_to_w}.  
In our calculations, we have assumed that the polar radius remains constant following \citet{collins1966}.
The stellar temperatures are defined using the assumption that the total luminosity of the star 
remains fixed which is a reasonable approximation (for discussion of this point, see \citet{lovekin2006}).   
These stellar parameters were chosen to include a model from each of the five bins adopted by \citet{cra05} to analyse 
the effects of spectral type on Be~star rotational statistics.  
We have calculated sets of models for the spectral types B0V, B2V, B3V, and B5V.  
Unfortunately, B8V models are too cool for {\sc Bedisk} when run for gravitational darkening 
near critical rotation without explicitly improving the treatment of the diffuse radiation field 
(i.e. disk self-heating) beyond the modified on-the-spot approximation used by \citet{sig07}

\vspace{-0.1in} 
\begin{table}[H]
\caption[]{Rotation rates. \label{v_to_w}}
 \vspace{-0.08in}
\begin{center}
\begin{tabular}{c c c c c c c}
\hline\hline
$v_{\rm{frac}}$& ${\omega}_{\rm{frac}}$ & ${r_{{\rm{eq}}}}/{r_{\rm{p}}}$ &\multicolumn{4}{c} {${v_{\rm{eq}}}$  (${\rm{km}}$ ${\rm{s}}^{-1} $) }       \\
               &              &          &B0V & B2V & B3V & B5V \\\hline
 0.0000 &  0.0000 & 1.00 &   0     &   0       &    0    & 0 \\
 0.2000 &  0.2960 & 1.01 &  110 &    93    &  89.7 & 87.7\\ 
 0.4000 &  0.5680 & 1.06 &  219 &   186   &  179  & 175\\ 
 0.6000 &  0.7920 & 1.14 &  329 &   280   &  269  & 263\\ 
 0.7000 &  0.8789 & 1.20 &  384 &   327   &  314  & 307\\
 0.8000 &  0.9440 & 1.27 &  439 &   373   &  359  & 351\\ 
 0.9000 &  0.9855 & 1.37 &  493 &   420   &  404  & 395\\
 0.9500 &  0.9963 & 1.43 &  521 &   443   &  426  & 417\\
 0.9900 &  0.9999 & 1.49 &  544 &   462   &  445  & 435\\ 
 \hline 
\end{tabular}
\end{center}
\end{table}

Models presented in \S~\ref{diff_density} and \S~\ref{diff_grav_dark} assumed a fixed density structure for the disk in the form, 
\begin{equation} 
\label{rho_disk}
\rho(R,z) = \rho_o \left( \frac{R_{*}}{R} \right)^{n} e^{-(z/H(R))^2} \, , 
\end{equation}
where 
\begin{equation} 
\label{scale_disk}
{H(R)} = \sqrt{\frac{2 R^3 k T_{\rm{iso}}}{G M \mu} } \, .  
\end{equation}
In Equation~(\ref{scale_disk}) for the disk scale height, 
$\mu$ is the mean-molecular weight of the gas (treated as a constant and set to half the mass of hydrogen) 
and $T_{\rm{iso}}$ is an assumed isothermal temperature used for the sole purpose of setting the density scale height; 
$T_{\rm{iso}}$ is set equal to 60\% of the effective temperature of the central star, listed in Table~\ref{star_para} \citep{hydro_paper}.  
The form of the disk density given by Equation~(\ref{rho_disk}) follows from the assumption of a radial power-law drop in the equatorial plane ($z=0$) coupled 
with the assumption that the disk is in vertical hydrostatic equilibrium set by the $z$-component of the star's gravitational acceleration 
and the assumed temperature $T_{\rm{iso}}$.  
This leads to a ``flaring disk" in which $H\propto\,R^{3/2}$.  

Models were made with the density fall off parameter, $n$, set to $3.0$.  
The disks were constructed with the unchanging grid described in \citet{me1} which keeps the disk empty for  $ r \le 1.5  \, r_{\rm{p}}$.  
This allows a disk of fixed density structure to be used for all calculations as the star expands with rotation up to a maximum of $1.5  \, r_{\rm{p}}$ at critical rotation.  
The filled region begins at $r=1.6 \, r_{\rm{p}}$ and the density in the mid-plane of the disk at the beginning of this filled region is given by 
$\rho_{\rm{bd}}= \rho_o \left( {1}/{1.6} \right)^3 =0.24 \rho_o$ using Equation~(\ref{rho_disk}).  
The parameters $\rho_o$ and $\rho_{\rm bd}$ will be referred to as the density scale and the base density respectively.  
The models were constructed for four different values of the density scale, 
$\rho_o$: $5.0\, \times \, 10^{-11}$; $2.5\, \times \, 10^{-11}$; $1.0\, \times 10^{-11}$; and $5.0\, \times \, 10^{-12}$ g cm$^{-3}$.  

Models presented in \S~\ref{hydro} have been hydrostatically converged which ensures that the vertical density structure is consistent 
with the temperature structure by enforcing hydrostatic equilibrium over an additional loop as described in \citet{hydro_paper}.  
Only eight hydrostatically converged models were made due to the long computation time required, $\approx10$ times longer than an unconverged model.  
Using the B0V stellar parameters, two hydrostatically converged models were calculated for each density scale, 
($\rho_o$: $5.0\, \times \, 10^{-11}$; $2.5\, \times \, 10^{-11}$; $1.0\, \times \, 10^{-11}$; and $5.0\, \times \, 10^{-12}$ g cm$^{-3}$ ).  
The first model without rotation and the second, with a rotation rate of $v=0.99\,v_{\rm{crit}}$.  

Models presented in \S~\ref{diff_grav_dark} use a version of {\sc Bedisk} that allows for variations in the formulation of gravitational darkening.  
Results for the B2V model were recalculated for the four fastest velocities in Table~\ref{v_to_w} at the four different density scales, $\rho_o $: 
$5.0\, \times \, 10^{-11}$; $2.5\, \times \, 10^{-11}$; $1.0\, \times \, 10^{-11}$; and $5.0\, \times \, 10^{-12}$ g cm$^{-3}$.  
Two variants of gravitational darkening were used: 
(1) reducing the value of the exponent on the magnitude of gravity in von Zeipel's theorem, $\beta$, from the standard value of $0.25$ to $0.18$;  
(2) changing the expression for the temperature structure based on \citet{von_Zeipel_1} to that given by \citet{Lara_2011}.  

\vspace{-0.1in}
\vspace{-0.1in}
\vspace{-0.1in}

\section{Results}
\label{results}
\vspace{-0.05in}
\subsection{Effects of Density Changes}
\label{diff_density}
\vspace{-0.05in}

\begin{figure}[H]
\epsscale{0.5}
\plotone{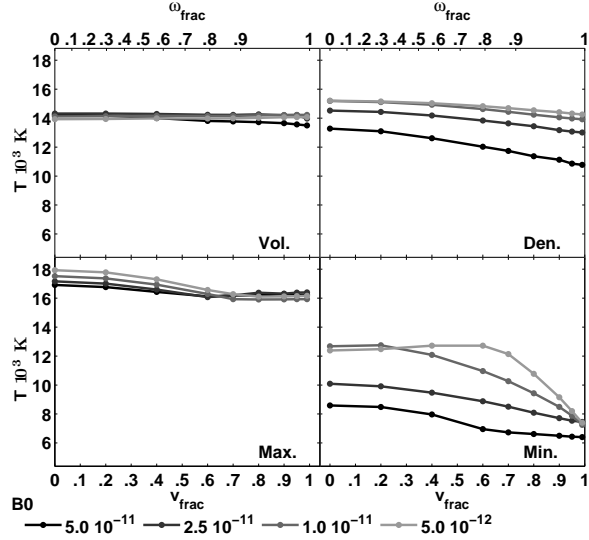}
\caption{Change in four disk temperature diagnostics with increasing stellar rotation 
and varying densities for the B0V model.  
Shown are 
the volume-weighted and the density-weighted average temperatures as 
defined in Equations~(\ref{T_mass_ave})~and~(\ref{T_vol_ave}) (upper left and upper right panels) and 
the maximum and minimum temperatures (lower left and lower right panels).  
Four different disk densities are shown with $\rho_o=$ 
$5.0\, \times \, 10^{-12}$; 
$1.0\, \times \, 10^{-11}$; 
$2.5\, \times \, 10^{-11}$; and 
$5.0\, \times \, 10^{-11}$ g cm$^{-3}$.  Darker lines indicate higher $\rho_o$.  
The lower horizontal axis indicates the fractional rotational velocity at the equator 
and the upper horizontal axis indicates the corresponding fractional angular velocity.  
\label{B0_max_min}}
\end{figure}

\begin{figure}[H]
\epsscale{0.5}
\plotone{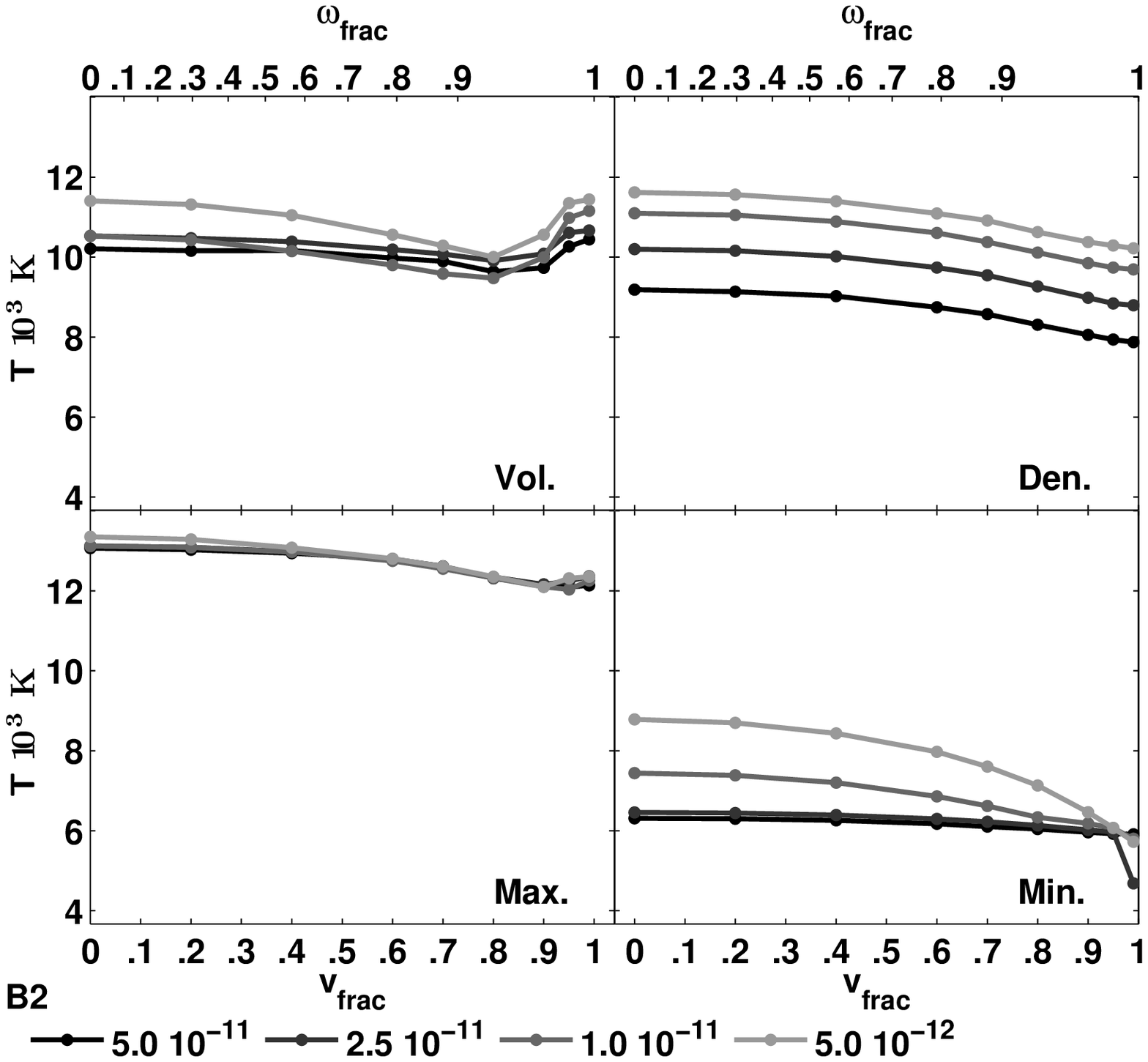}
\caption{Same as Figure~\ref{B0_max_min} but for the B2V model. \label{B2_max_min}}
\end{figure}

\begin{figure}[H]
\epsscale{0.5}
\plotone{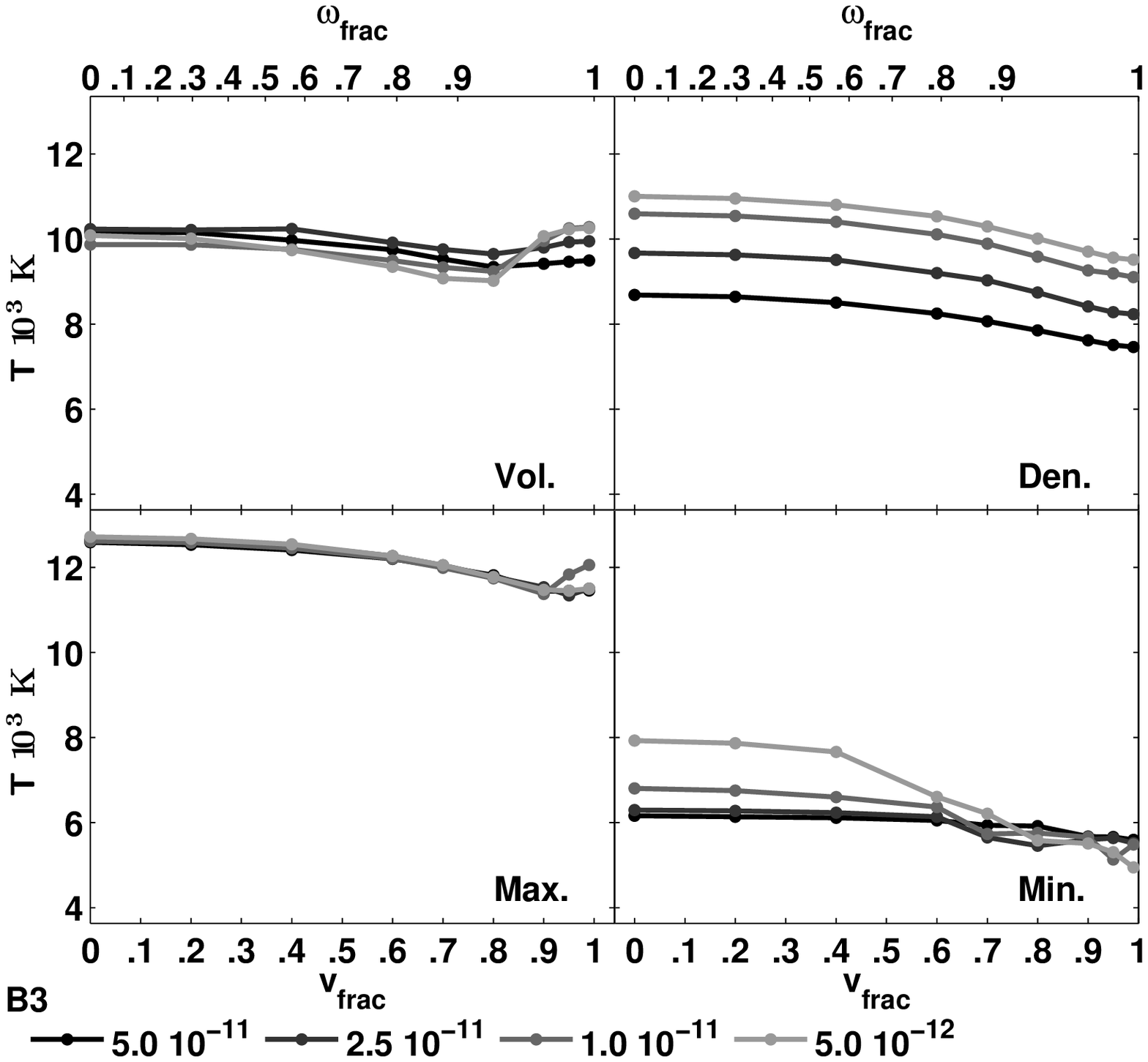}
\caption{Same as Figure~\ref{B0_max_min} but for the B3V model. \label{B3_max_min}}
\end{figure}

\begin{figure}[H]
\epsscale{0.5}
\plotone{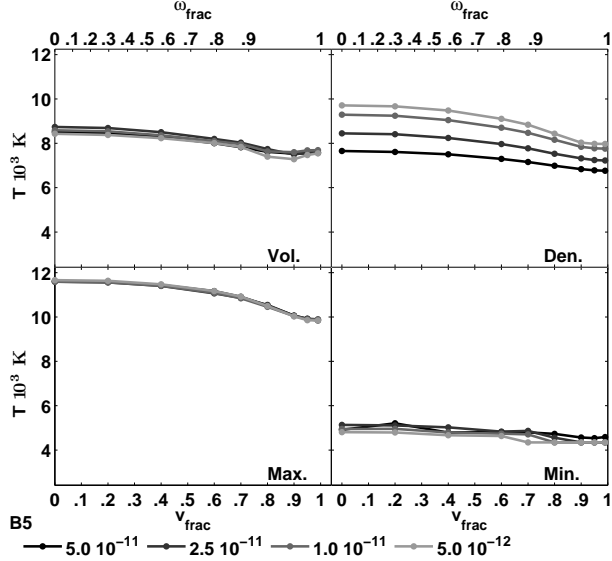}
\caption{Same as Figure~\ref{B0_max_min} but for the B5V model. \label{B5_max_min}}
\end{figure}
Figures~\ref{B0_max_min}, \ref{B2_max_min}, \ref{B3_max_min}, and~\ref{B5_max_min} show the changes produced by increasing rotation 
on the temperature of disks for a variety of densities using the same four temperature 
diagnostics as \citet{me1}: the density-weighted average temperature, the volume-weighted average temperature, 
the maximum temperature, and the minimum temperature.  
The density-weighted average temperature is defined as 
\begin{equation}
\vspace{-0.05in}
\label{T_mass_ave}
\overline{T}_{\rho} = \frac{1}{M_{\rm{disk}}}\,\int T(R,z)\,\rho(R,z)\,dV \; , 
\vspace{-0.05in}
\end{equation}
and the volume-weighted average temperature is defined as 
\begin{equation}
\vspace{-0.05in}
\label{T_vol_ave}
\overline{T}_{\rm{V}} = \frac{1}{V_{\rm{disk}}}\,\int T(R,z)\,dV \;.  
\vspace{-0.05in}
\end{equation}
In order to avoid numerical effects, the median of the twenty hottest and 
twenty coolest disk locations are taken to represent the maximum and minimum disk temperatures.

The most important trend seen in Figures~\ref{B0_max_min} through \ref{B5_max_min} is in the density-weighted average temperatures.  
There is a decrease in this temperature with both increasing rotation rate and increasing density.  
This trend is seen in all spectral types.  
In addition, for all spectral types, the curves for different $\rho_o$ are well separated and decrease smoothly.  
For spectral types B2V, B3V, and B5V, the slope increases with increasing rotation but flattens out near critical rotation.  
For spectral types B2V and B3V, the four curves for each $\rho_o$ are essentially parallel and for B5V are nearly parallel.  
Parallel curves are not seen in the B0V model, and the decrease in temperature with rotation is greatest for the largest density model.  

The temperature minimums also decrease with both increasing rotation rate and 
increasing density, analogous to the density-weighted average temperatures.  
The size of the decrease in the minimum temperature is also larger for the least dense models.  
For spectral types B2V, B3V, and B5V, the curve for the the most dense model, $\rho_o=5.0\, \times \, 10^{-11}$ g cm$^{-3}$, 
is essentially flat.  
This causes the temperature minimums to converge at $v_{\rm{crit}}$ to values between 5000 and 6000 K for all spectral types.  
The curves for different $\rho_o$ are distinct, but there is significant overlap.  
The temperature minimums in the non-rotating models are larger for the earlier spectral types, 
and the earlier models experience a larger decrease in temperature with both rotation and density.  
The drop in the value of the minimum temperatures with rotation and density is barely noticeable in the B5V models 
as all curves are nearly flat and cluster at $\approx$5000 K.  

One of the most interesting effects produced by rotation on Be~star disk temperatures is seen in 
the volume-weighted average temperatures of the B2V, B3V, and B5V models.  
These values initially decrease with moderate rotation but begin to increase approaching critical rotation.  
The strength of both the initial decease and the increase near critical rotation is larger for early spectral types, excluding B0V.  
For the two highest densities of the critically rotating B2V models, 
the volume-weighted average temperatures are the essentially the same as the non-rotating models.  
Generally the volume-weighted average temperatures are hotter for the least dense models, 
but the separations between the averages are not as large as for the density-weighted average temperatures 
or the temperature minimums, and there is a great deal of overlap.  
The effect of density is strongest in the B2V model and very small by B5V.  
For moderate rotation, lower density models generally have smaller volume-weighted temperature averages than denser models, 
but the lowest density B2V models ($\rho=5.0 \, \times \, 10^{-12}$), 
and the non-rotating and slowest rotating models of the second lowest density B2V models 
($\rho=1.0 \, \times \, 10^{-11}$) do not follow this trend.  
For rapid rotation, lower density models generally have larger volume-weighted temperature averages than denser models, 
but this does not occur for the B5V models.    

The volume-weighted temperature averages of the B0V model does not behave like the other spectral types.  
There is only a very small change in temperature with rotation, which is somewhat larger for larger densities.  
The least dense B0V model is of nearly constant temperature.  
The densest, non-rotating models are hotter than the least dense models, 
but by critical rotation, this trend has reversed.  
All of the values are tightly clustered between 13900 to 14500 K.  

The maximum disk temperatures change very little with rotation or density.  
This is because these temperatures are found in parts of the disk that are not shielded by the dense, 
mid-plane and are directly illuminated by the pole of the star.  
The maximum temperatures are determined essentially by spectral type.  
There is a small but noticeable decrease in temperature with rotation in all models, which flattens out near critical rotation.  
Like the volume-weighted average temperatures, 
the temperature maximums begin to increase for extreme rotation, 
significantly for the B3V models but very weakly for the other spectral types.  

One of the most complex behaviours seen in Figures~\ref{B0_max_min} through \ref{B5_max_min} 
is the how the relationship between the volume-weighted and 
the density-weighted average temperatures is affected by rotation and disk density.  
These effects can be separated into three categories based on density, the high and low density extremes and the case of intermediate density.  
(1) For the highest density B0V model, 
the two highest density B2V and B3V models, 
and the highest density B5V model, 
the volume-weighted average temperatures are always higher than the density-weighted average temperatures.  
This is due to the presence of a large, cool region in the equatorial plane at high density.  
(2) For the lowest density B0V models and the two lowest density B5V models, 
the volume-weighted average temperatures are always lower than the density-weighted average temperatures.  
This is due to the absence of a significant cool region in the equatorial plane at low density.  
(3) All models of intermediate density experience a transition: 
for low rotation rates there is either not a cool region in the equatorial plane or 
it is too small to cause the density-weighted average temperatures to be lower then the volume-weighted average temperatures.  
At higher rotation rates, gravitational darkening causes the development of a larger cool region 
and the density-weighted average temperatures become less than the volume-weighted average temperatures.  

\begin{figure}[H]
\vspace{-0.1in}
\epsscale{0.5}
\plotone{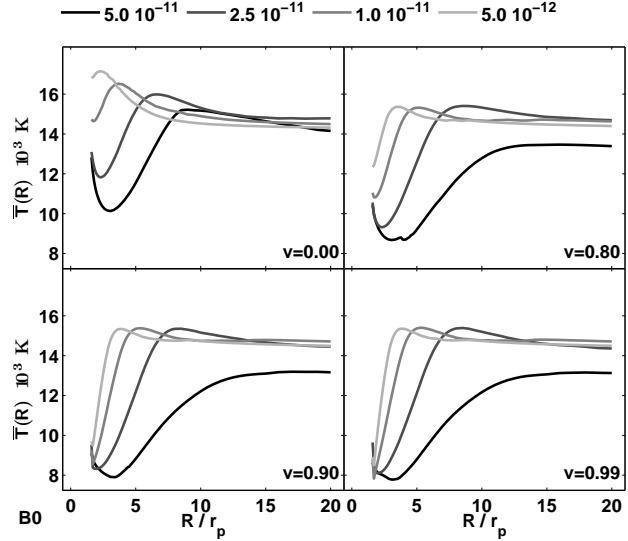}
\vspace{-0.2in}
\caption{Variation of the vertically-averaged, density-weighted average temperature with 
disk radius for various stellar rotation rates and disk densities for the B0V model.  
Each panel shows a different rotation rate: no rotation (upper left); 
$v=0.80\,v_{\rm{crit}}$ (upper right); 
$v=0.90\,v_{\rm{crit}}$ (lower left); and $v=0.99\,v_{\rm{crit}}$ (lower right).  
Four different disk densities are shown for each rotation rate: with $\rho_o=$ 
$5.0\, \times \, 10^{-12}$; 
$1.0\, \times \, 10^{-11}$; 
$2.5 \, \times \, 10^{-11}$; and 
$5.0 \, \times \, 10^{-11}$ g cm$^{-3}$.  
The darker lines indicate higher $\rho_o$.  
\label{B0_rprofile}}
\vspace{-0.1in}
\end{figure}

\begin{figure}[H]
\epsscale{0.5}
\plotone{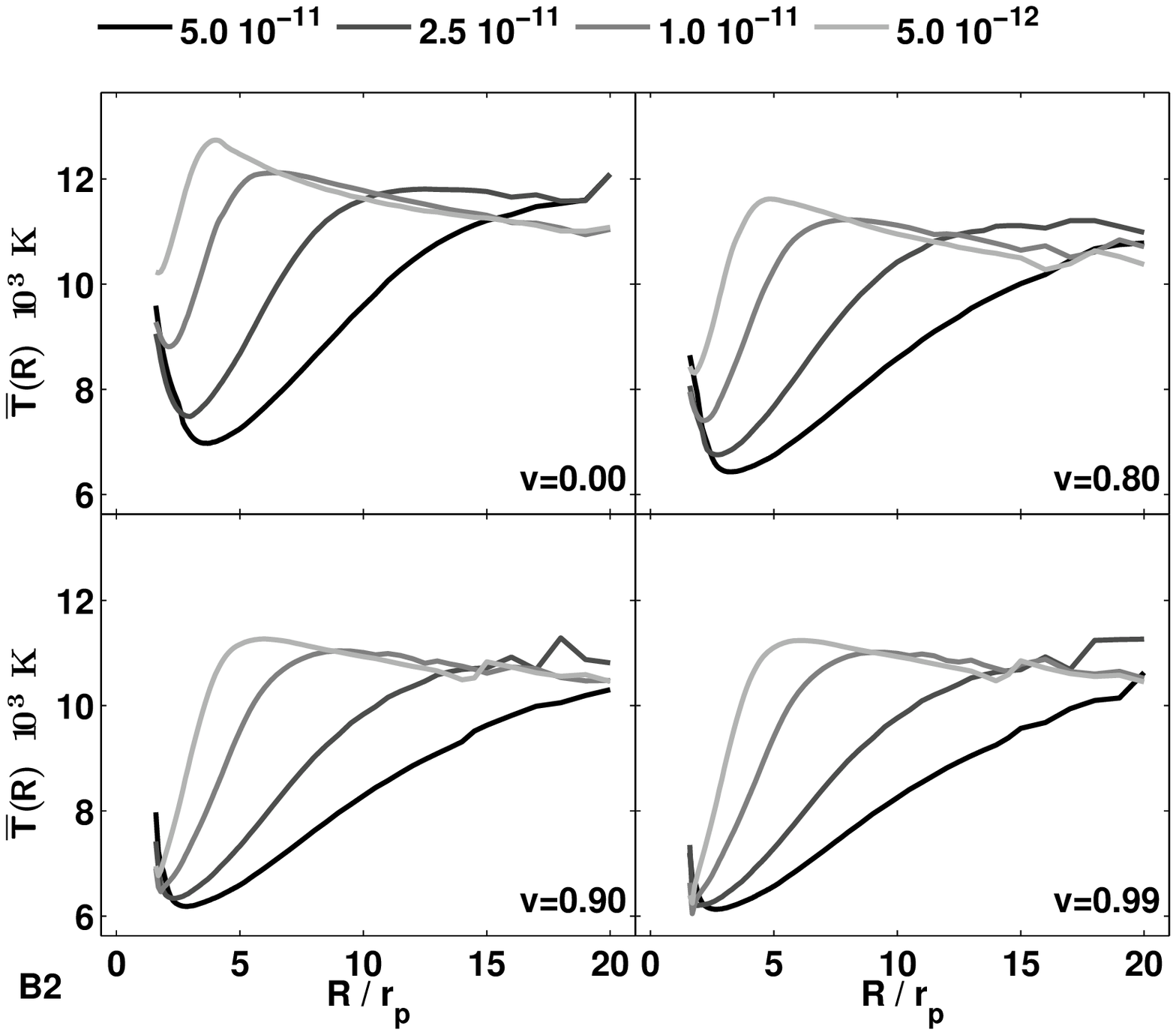}
\caption{Same as Figure~\ref{B0_rprofile} for the B2V model.  \label{B2_rprofile}}
\end{figure}

\begin{figure}[H]
\epsscale{0.5}
\plotone{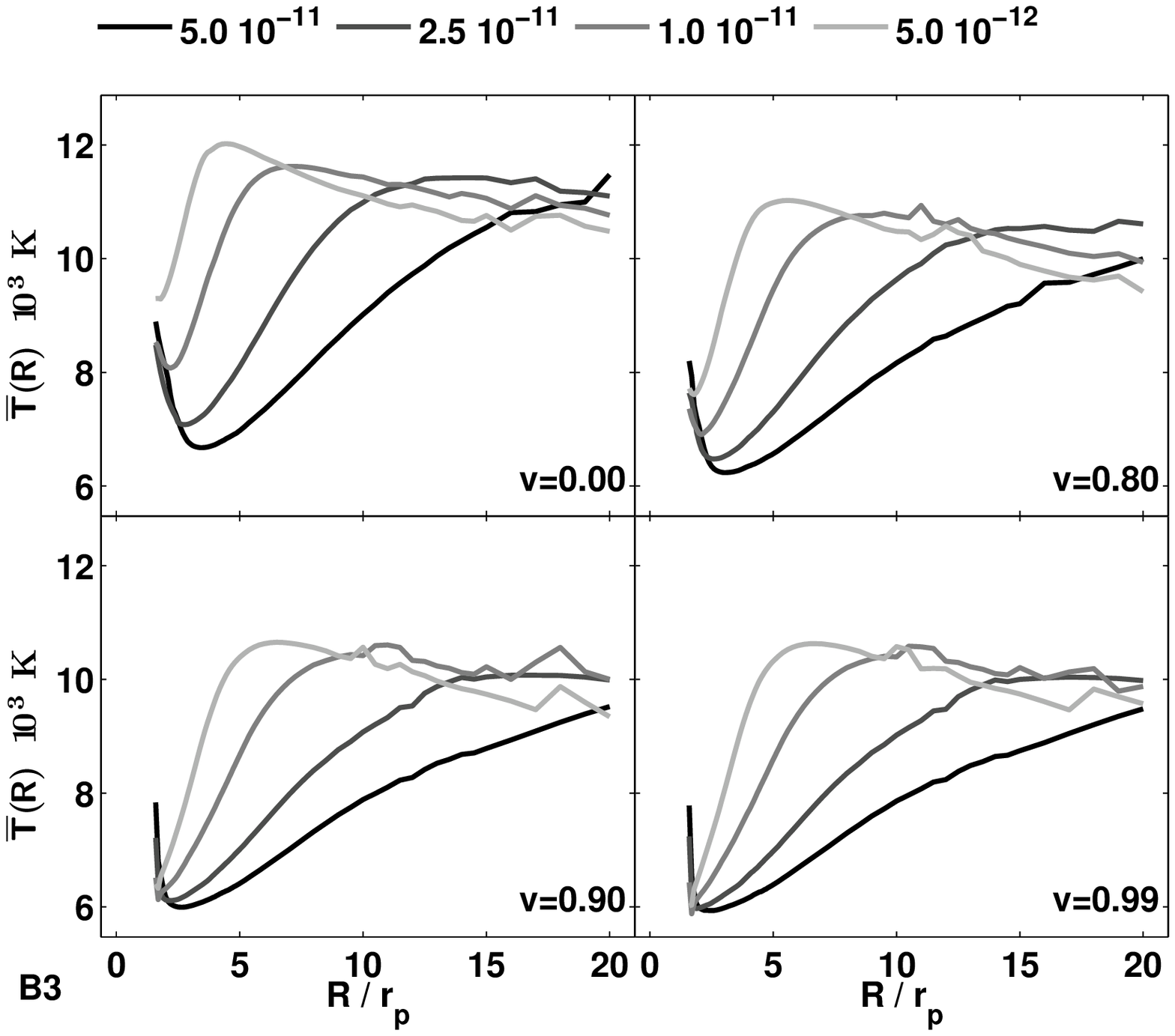}
\caption{Same as Figure~\ref{B0_rprofile} for the B3V model.  \label{B3_rprofile}}
\end{figure}

\begin{figure}[H]
\epsscale{0.5}
\plotone{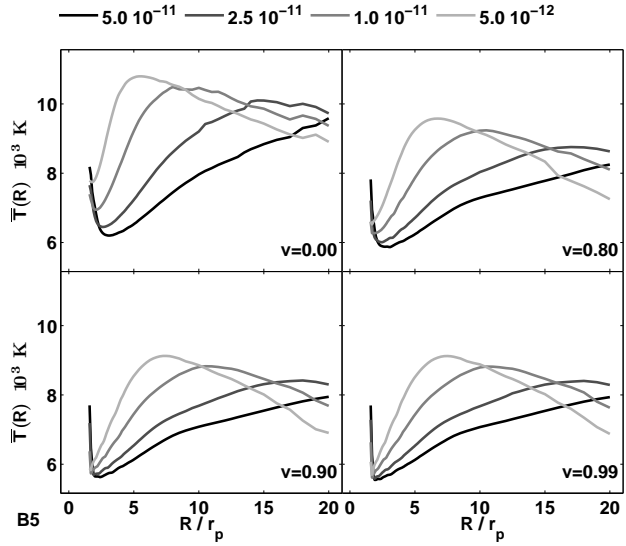}
\caption{Same as Figure~\ref{B0_rprofile} for the B5V model.  \label{B5_rprofile}}
\end{figure}

Figures~\ref{B0_rprofile}, \ref{B2_rprofile}, \ref{B3_rprofile}, and~\ref{B5_rprofile} show radial temperature profiles of these disks 
obtained by averaging the temperatures within each vertical column perpendicular to the mid-plane of the disk.  
The average temperature at each radial distance is density-weighted and is found using 
\vspace{-0.125in}
\begin{equation}
\vspace{-0.19in}
\label{eq_r_profile}
\bar{T}(R)=\frac{\int^{z_{\rm {max}}}_{0} \, T(R,z) \rho (R,z) dz}{\int^{z_{\rm{max}}}_{0} \rho (R,z) dz} \, .  
\end{equation}
\vspace{-0.05in}

As seen in the previous figures, lower density models have higher temperatures.  
The profiles follow two patterns: 
(1) for low density models,
the temperature at the inner edge of the disk begins fairly hot, increases to a maximum
and then the temperature decreases as the radius increases, sometimes reaching a constant value; 
(2) for sufficiently large densities, the temperature at the front of the disk still begins fairly hot,
but the temperature drops to a minimum and then increases again at larger radii,
sometimes reaching a maximum value and then dropping again.  
For all spectral types, the temperature minimum becomes cooler and the extent of the cool zone increases in size with rotation.  
The slope of the temperature increase at large radii becomes smaller with increasing rotation.  
If a temperature maximum occurs, 
its size is reduced with rotation and the slope of the temperature drop off at large radii becomes shallower.

\begin{figure}[H]
\epsscale{0.5}
\plotone{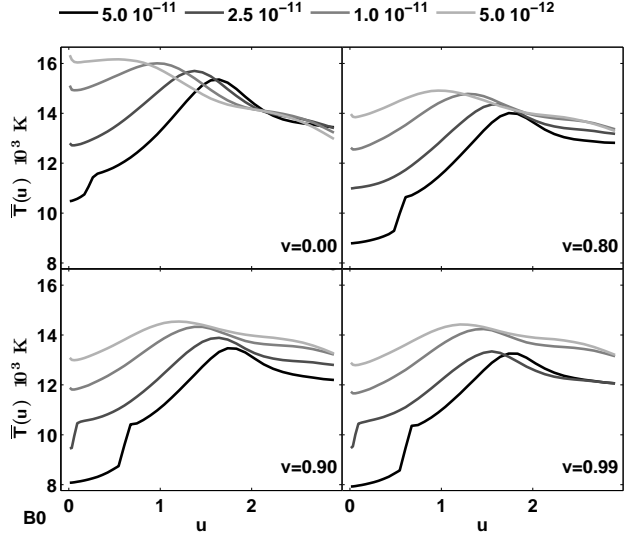}
\caption{ Variation of the radially-averaged temperature profile 
for various stellar rotation rates and disk densities for the B0V model.  
Here the temperature is plotted versus the scale height, u, defined by $u \equiv z/H(R)$.  
Each panel shows a different rotation rate: no rotation (upper left); 
$v=0.80\,v_{\rm{crit}}$ (upper right); 
$v=0.90\,v_{\rm{crit}}$ (lower left); and $v=0.99\,v_{\rm{crit}}$ (lower right).  
Four different disk densities are shown for each rotation rate with $\rho_o=$ 
$5.0\, \times \, 10^{-12}$; 
$1.0\, \times \, 10^{-11}$; 
$2.5\, \times \, 10^{-11}$; and 
$5.0\, \times \, 10^{-11}$~g cm$^{-3}$.  
Darker lines indicate higher $\rho_o$.  
\label{B0_zprofile}}
\end{figure}

\begin{figure}[H]
\epsscale{0.5}
\plotone{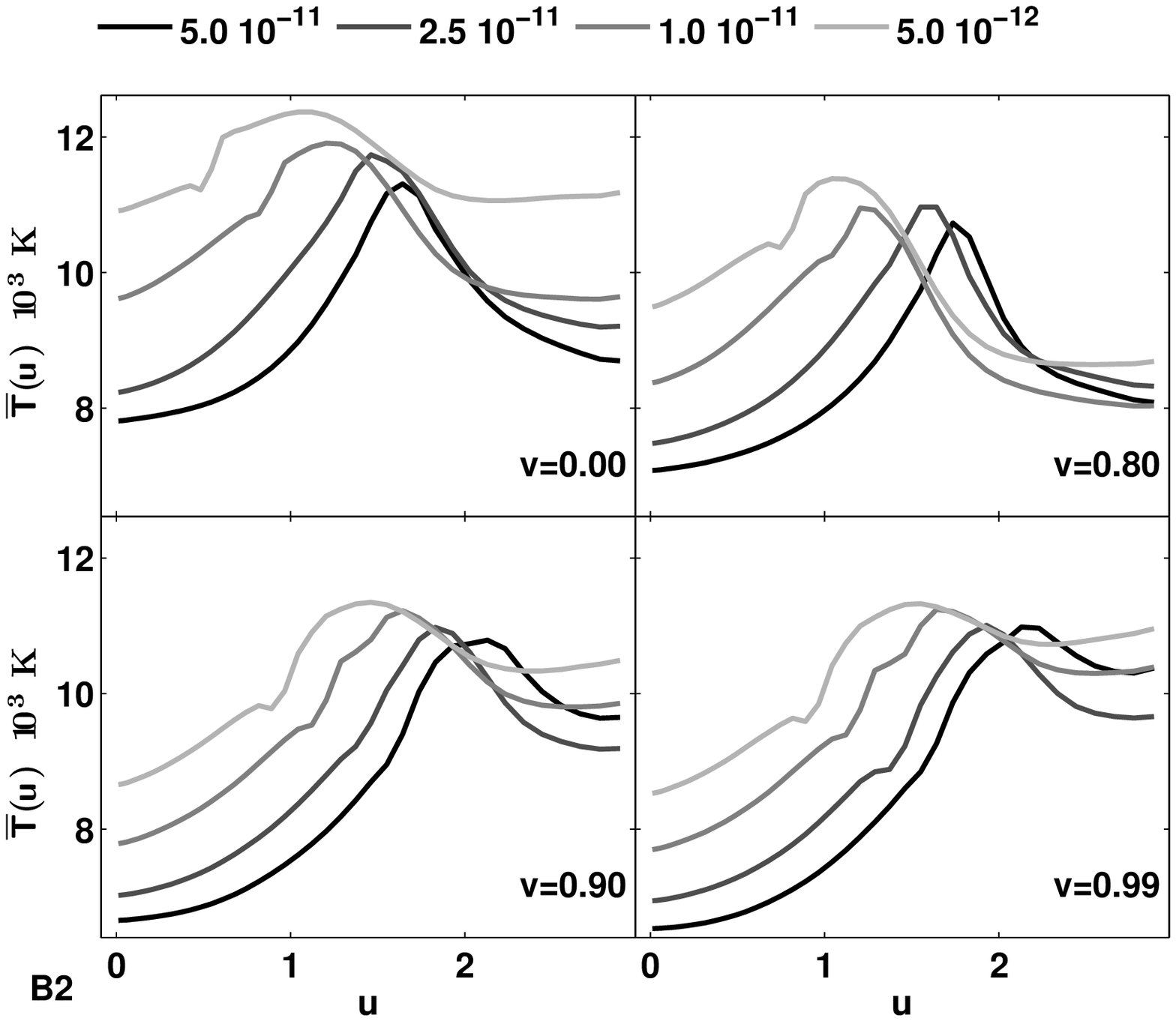}
\caption{Same as Figure~\ref{B0_zprofile} for the B2V model.  \label{B2_zprofile}}
\end{figure}

\begin{figure}[H]
\epsscale{0.5}
\plotone{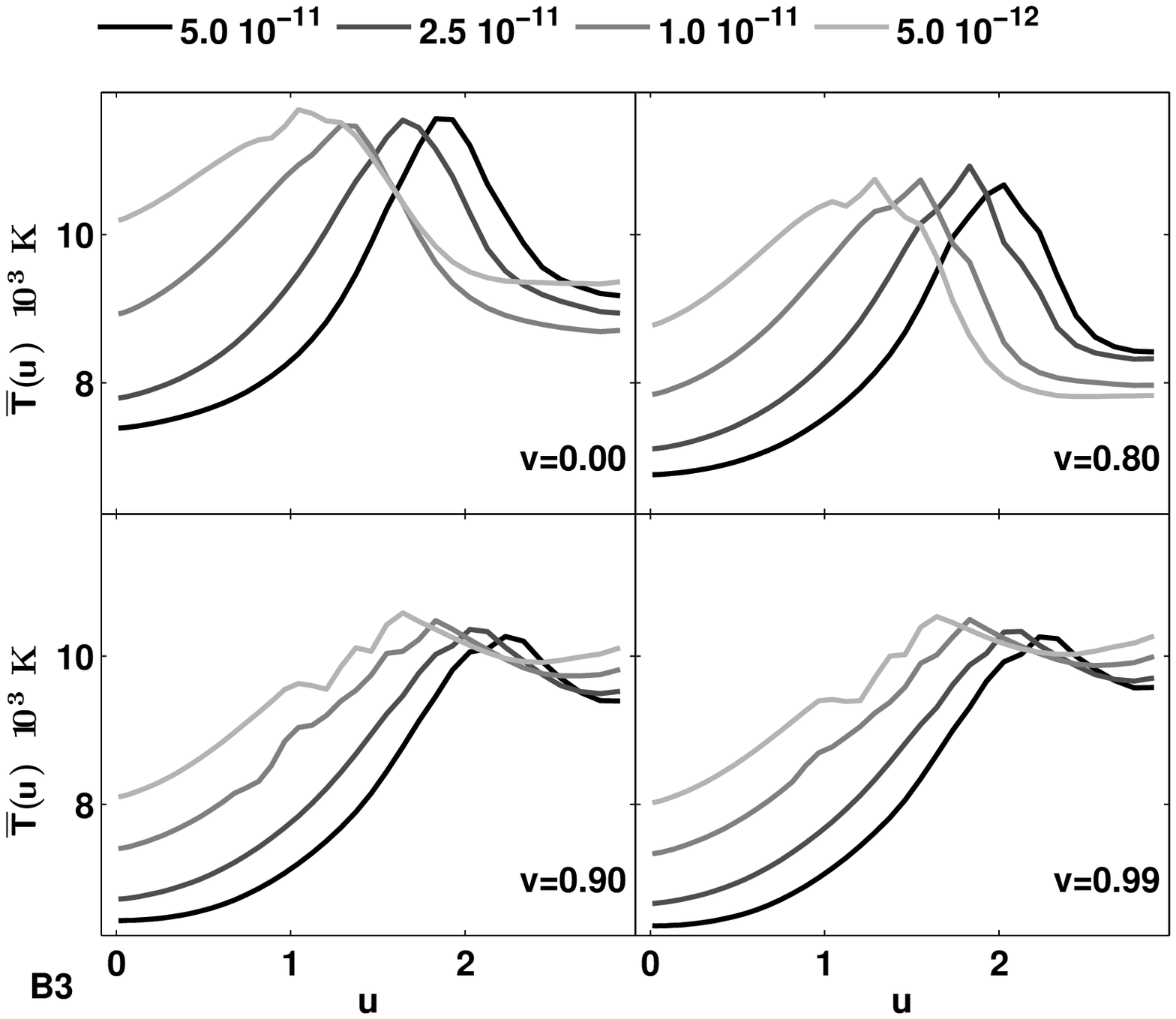}
\caption{Same as Figure~\ref{B0_zprofile} for the B3V model.  \label{B3_zprofile}}
\end{figure}

\begin{figure}[H]
\epsscale{0.5}
\plotone{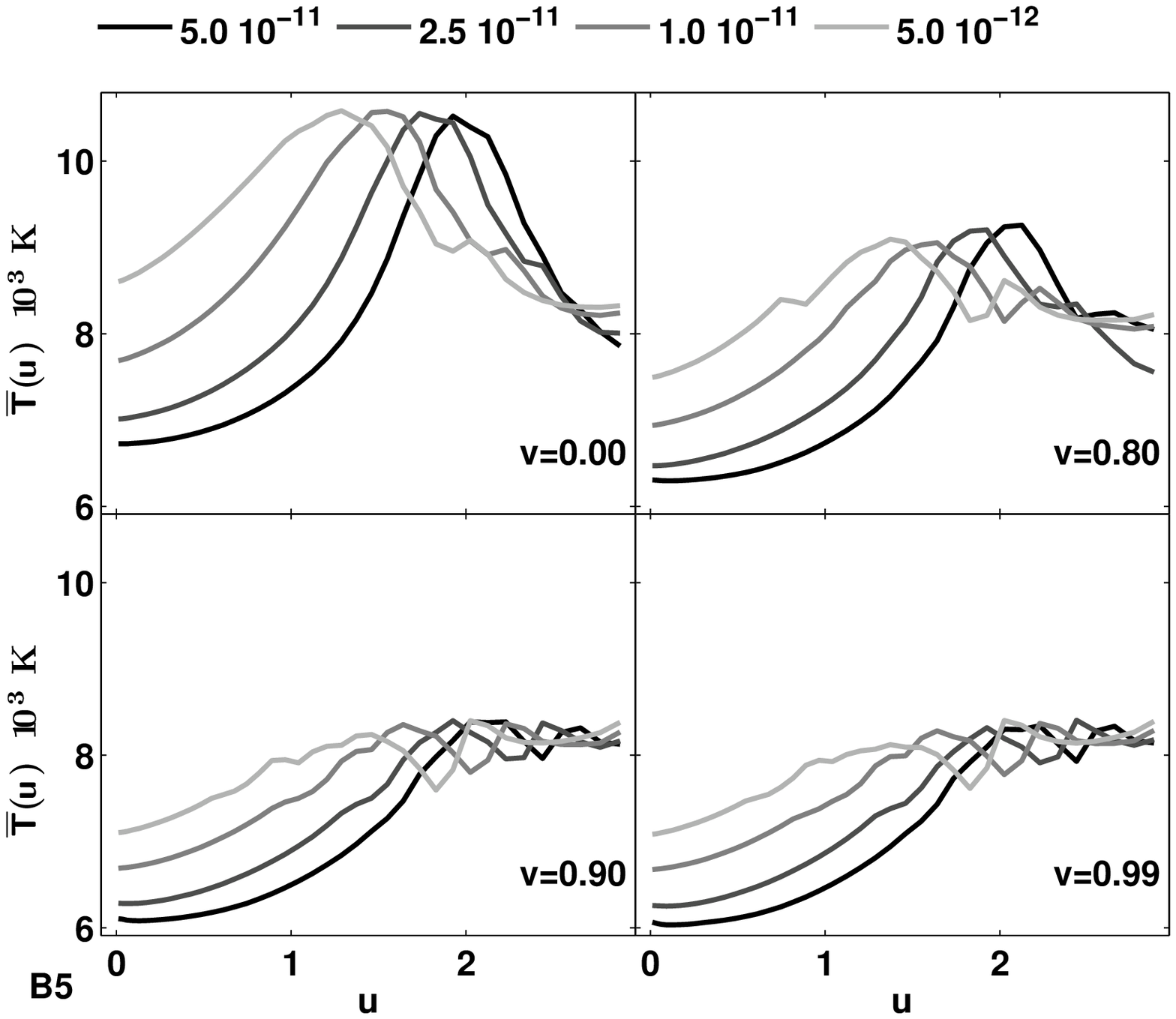}
\caption{Same as Figure~\ref{B0_zprofile} for the B5V model.  \label{B5_zprofile}}
\end{figure}

Figures~\ref{B0_zprofile}, \ref{B2_zprofile}, \ref{B3_zprofile}, and~\ref{B5_zprofile} show vertical temperature 
profiles of these disks obtained by averaging the temperatures at different radii but 
at the same scale height, $u=z/H(R)$.  
The average temperature is density-weighted and is found using 
\begin{equation}
\label{eq_u_profile}
\bar{T}(u)=\frac{\int^{R_{\rm{ max}}}_{0} \, T(R,u) \rho (R,u) A(R,u) dR}{\int^{R_{\rm{ max}}}_{0} \rho (R,u) A(R,u) dR} \, , 
\end{equation}
where $A(R,u)$ is the area function of the disk.  $A(R,u)$ is included in Equation~\ref{eq_u_profile} 
to account for the non-uniform spacing in $r$ and $z$ of the computational grid (i.e. $\Delta r$ and $\Delta z$ both increase with R).  
The exact nature of this weighting matters less than the fact that all the models have been averaged in the same way.  
Comparing these plots illustrates the effects of rotation and density on the vertical structure of these disks.  

In these plots, the mid-plane of the disk is at u=0.  
All vertical temperature profiles show a temperature maximum occurring between one and two scale heights above the mid-plane.  
The location of this maximum moves higher above the mid-plane with both increasing density and increasing rotation.  
Rotation causes the width of the temperature peak to increase and become broader, 
while increasing density causes width of the peak to become narrower and sharper.  
Low density models without rotation have mid-plane regions either at the same temperature or hotter than the upper disk.  
Increasing the density causes the development of a cool region in the mid-plane.  
Rotation causes this cool region to form in the mid-plane at lower densities.  
At high rotation rates, pronounced cool regions are seen even in the least dense models.  

There is little variation in the temperature of the upper edge of the disk with either rotation or density in the B5V models 
and no significant change with density for the non-rotating B0V models.  
The upper edges of the disk increase in temperature with rapid rotation in the B2V and B3V models and have a minimum value at $v_{\rm {frac}}=0.80$.  
In the B2V, B3V and the rotating B0V models, the upper edges of the disk are hotter for lower densities.  
While there is a change in the temperature of the upper edges of the disk with density in the rotating B0V models, 
which increases in size with rotation, the median of this range is not significantly affected by rotation and remains at $\approx 13000$ K.  

\begin{figure}[H]
\epsscale{0.87}
\plotone{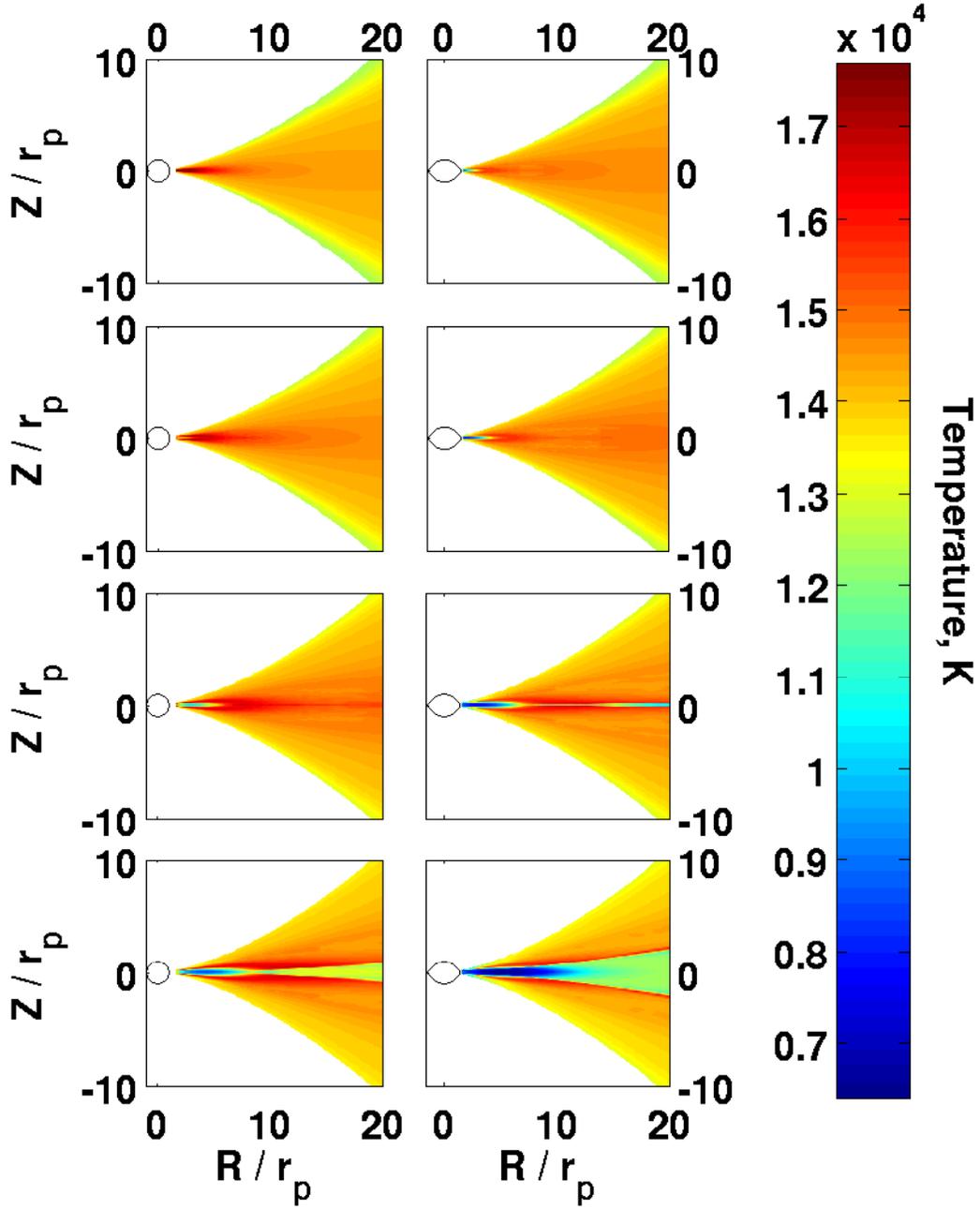}
\caption{Temperature profiles of the disk for a selection of B0V models.  
The left column shows models with no rotation.  
The right column shows models with a rotational speed of $v_{\rm{frac}}=0.95$.  
The top row shows models with $\rho_o=5.0\, \times \, 10^{-12}$ g cm$^{-3}$; 
the second row, $\rho_o=1.0\, \times \, 10^{-11}$ g cm$^{-3}$; 
the third row, $\rho_o=2.5\, \times \, 10^{-11}$ g cm$^{-3}$; 
and the bottom row, $\rho_o=5.0\, \times \, 10^{-11}$ g cm$^{-3}$.  
The colour-map indicates the disk temperatures.  
The black line outlines the star.  \label{2D_panel_B0}}
\end{figure}

Figure~\ref{2D_panel_B0} shows the two dimensional temperature structure for a selection of eight B0V models.  
Two models are shown for each density, one without rotation and one for $v_{\rm{frac}}=0.95\,v_{\rm{crit}}$.  
Clearly the temperature in the equatorial plane becomes cooler as the density increases; this has been noted many times.  
What is interesting is that this cool region appears at smaller densities when rotation is included.  

\begin{figure}[H]
\epsscale{0.5}
\plotone{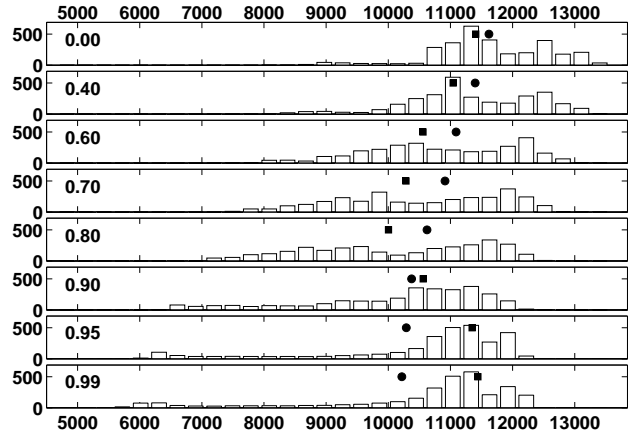}
\caption{Series of histograms of disk temperatures.  
All histograms plotted are for the B2V model with $\rho_o=5.0 \, \times \,  10^{-12}$ g cm$^{-3}$.  
Each panel shows a model with a different velocity as indicated in the left of each panel.  
The upper most panel is the non-rotating model and the rotation rates increase downward.  
The lower most panel is the fastest rotating model with $v_{\rm{frac}}=0.99 \, v_{\rm{crit}} $.  
The filled circles indicate the density-weighted average temperatures 
and the filled squares indicate the volume-weighted average temperatures for each model.  
\label{hish_5d0em12}}
\end{figure}

\begin{figure}[H]
\epsscale{0.5}
\plotone{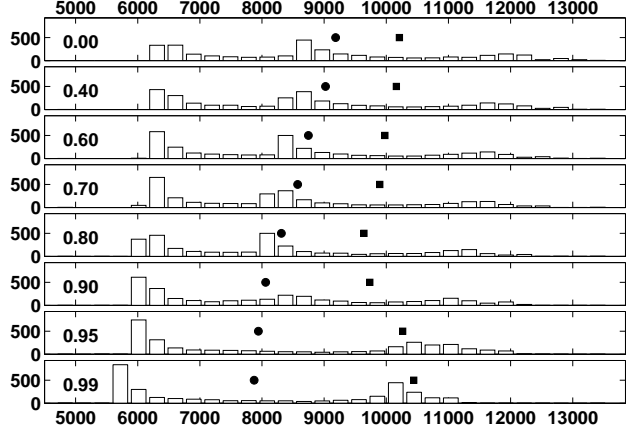}
\caption{Same as Figure~\ref{hish_5d0em12} but with $\rho_o=5.0 \, \times \, 10^{-11} $ g cm$^{-3}$. \label{hish_5d0em11}}
\end{figure}

Finally a useful way to illustrate the effect of density and rotation 
on the range of disk temperature is to construct histograms of $T(R,z)$ for each model.  
Figures~\ref{hish_5d0em12} and~\ref{hish_5d0em11} each show a set of histograms for increasing rotation.  
Only two sets of histograms are shown for brevity, 
the B2V model with $\rho_o=5.0 \, \times \, 10^{-12} $ g~cm$^{-3}$ (the lowest density considered) 
and another with $\rho_o=5.0 \, \times \, 10^{-11} $ g~cm$^{-3}$ (the highest density considered).  
In Figure~\ref{hish_5d0em12}, the model without rotation is fairly warm and as rotation increases, a low temperature tail forms in the distribution.  
In addition, the fraction of disk temperatures in the highest temperature bins also increases at high rotation rates.  
In Figure~\ref{hish_5d0em11} the non-rotating model has a two strong peaks, one quite cool and a second middle peak.  
There is also a large and long high temperature tail 
and a weak high temperature peak because this model is dense enough to possess a cool region in the equatorial plane, with warmer regions above.  
As rotation increases, the middle temperature peak weakens and eventually disappears, while the low temperature peak becomes stronger and cooler.  
The weak, higher temperature peak becomes stronger.  
The distribution of temperatures in the fastest rotating model has two strong peaks at the temperature extremes.  
Overall, these histograms demonstrate that gravitational darkening increases the amount of very cool and very hot material in the disk 
and decreases the amount of disk material of intermediate temperatures.  

\begin{figure}[H]
\epsscale{0.5}
\plotone{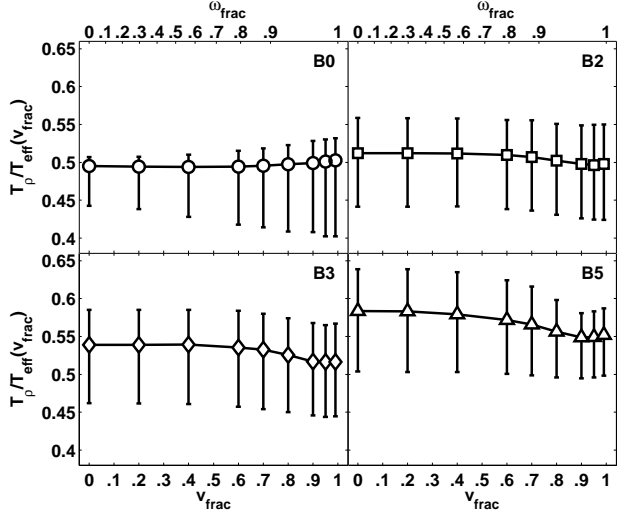}
\caption{
Range of density-weighted average temperatures for the all model disks divided by a measure of stellar effective temperatures 
including the effects of rotation (see Equation~(\ref{T_eff_rotation})) versus rotation.  
The median temperature for the four different densities is plotted.  
The upper error bar corresponds to the highest value (and the lowest density model) and 
the lower error bar corresponds to the lowest value (and the highest density model).  
Each panel is a spectral type as indicated.  
The lower horizontal axis indicates the fractional rotational velocity at the equator 
and the upper horizontal axis indicates the corresponding fractional angular velocity.  
\label{last_plot_2}}
\end{figure}

It is often useful in the study of circumstellar material to compare the temperatures found in the disk to the effective temperature of the star.  
However, when the star is a rapid rotator, there is a range of surface temperatures, and some representative value must be found.  
The most basic definition of effective temperature is from the Stephan-Boltzmann law, 
namely $T_{\rm{eff}} \equiv \left( L/\sigma A \right)^{1/4}$ where $A$ is the surface area of the star.  
Extending this to a rotating star gives, $T_{\rm{eff}}(v_{\rm{frac}})= \left( {L}/{\sigma A(v_{\rm{frac}})} \right)^{1/4}$.  
If the luminosity is considered unaffected by rotation we have, 
\begin{equation}
\label{T_eff_rotation}
T_{\rm{eff}}(v_{\rm{frac}})={T_{\rm{eff}}(v_{\rm{frac}}=0)} \left( \frac{A(v_{\rm{frac}}=0)}{A(v_{\rm{frac}})}\right)^{1/4} \, .  
\end{equation}
Figure~\ref{last_plot_2} shows the density-weighted average temperatures divided by the stellar effective temperatures 
defined by Equation~(\ref{T_eff_rotation}) versus rotation rate for all models.  
This plot summarizes the predicted global temperatures of the disks around rapidly rotating stars of a wide range of disk densities.  
The temperature ratio of the disk to the star is between $0.40$ - $0.65$ for all models.  
Denser disks are proportionally cooler.  
The disks around cooler stars all have temperatures 
which are larger fractions the stellar effective temperatures than those around hot stars.  
The ratio between the density-weighted average temperatures in the disk and the effective stellar temperatures drops with moderate rotation.  
Approach to critical rotation causes an increase in the ratios with rotation in both cooler stars and less dense disks.  
The slope of the ratio flattens out for rapid rotation in hot and dense disks.  
In all cases, the trends are dominated by the variations in disk density, not rotation.  

Table~\ref{density_table} gives the density-weighted temperatures for all density scales and 
spectral types without rotation and at maximum rotation ($v=0.99\, v_{\rm{crit}}$).  
This table shows the full range of temperature changes that occur with rotation for different densities.  
Increases in either density or rotation cause decreases in the density-weighted temperatures.  
For the B0V models, these two effects strengthen each other; denser disks are more strongly affected by increasing rotation and 
 disks surrounding critically rotating stars are more strongly affected by density changes than disks surrounding non-rotating stars.  
The density-weighted average temperatures change by $-6$ \% as rotation varies from zero to $0.99 \, v_{\rm{crit}}$ for the least dense models, 
and change by $-21$ \% for the densest model over the same range in velocity.  
When we look at the results with rotation rates fixed, we find that 
the density-weighted average temperatures change by 
$-14$ \% from $\rho_o=5.0\, \times \, 10^{-12}$ g cm$^{-3}$ to $\rho_o=5.0\, \times \, 10^{-11}$ g cm$^{-3}$ without rotation, 
and change by $-28$ \% for the fastest rotating models.  
In the B2V and B3V models, the effect of rotation is nearly the same for all $\rho_o$, 
and the percent differences range between $-13$ and $-15$ \% for the B2V models 
and between $-15$ and $-16$ \% for the B3V models.  
The effect of density changes is also nearly same on models without rotation 
and those with a rotation rate of $0.99\,v_{\rm crit}$, with 
the percent differences ranging between $-23$ and $-26$ \% 
for the B2V models and remaining unchanged at $-24$ \% for the B3V models. 
By B5V the these two effects weaken each other.  
The least dense models are more strongly affected by rotation 
and the non-rotating models are most affected by density changes.  
The density-weighted average temperatures change by $-20$ \% from zero to $0.99 \, v_{\rm{crit}}$ for the least dense models, 
and change by $-12$ \% for the densest model.  
Holding rotation rates fixed, we find that the density-weighted average temperatures change by 
$-24$ \% from $\rho_o=5.0\, \times \, 10^{-12}$ g cm$^{-3}$ to $\rho_o=5.0\, \times \, 10^{-11}$ g cm$^{-3}$ without rotation, 
and change by $-16$ \% for the fastest rotating models.  

 \vspace{-0.08in}
\begin{table}[H]
\caption[]{Density-weighted average temperatures without rotation and with $v_{\rm{frac}}=0.99$ \label{density_table}}
 \vspace{-0.08in}
\begin{center}
\begin{tabular}{c c c c c}
\hline \hline
Type&               $\rho_o$                 &$\bar{T}_{\rho}(0.00)$&$\bar{T}_{\rho}(0.99)$& \% $\Delta (v_{\rm frac})$  \\  
         &                 g cm$^{-3}$           &                  K                       &                  K                        &                    \\  \hline 
 B0V&$5.0 \, \times \, 10^{-12}$&15200.&14250.& - 6.\\
         &$1.0 \, \times \, 10^{-11}$&15180.&13920.& - 9.\\
         &$2.5 \, \times \, 10^{-11}$&14520.&13010.& -11.\\
         &$5.0 \, \times \, 10^{-11}$&13280.&10780.& -21.\\
         &{\% $\Delta (\rho)$}          &    -14. &     -28.&        \\ \hline 
 B2V&$5.0 \, \times \, 10^{-12}$&11620.&10220.& -13.\\ 
         &$1.0 \, \times \, 10^{-11}$&11090.&  9690.& -13.\\
         &$2.5 \, \times \, 10^{-11}$&10200.&  8800.& -15.\\
         &$5.0 \, \times \, 10^{-11}$&  9190.&  7870.&- 15.\\
         &{\% $\Delta (\rho)$}          &    -23. &     -26.&        \\ \hline 
 B3V&$5.0 \, \times \, 10^{-12}$&11000.&  9520.& -15.\\
         &$1.0 \, \times \, 10^{-11}$&10590.&  9110.& -15.\\
         &$2.5 \, \times \, 10^{-11}$&  9670.&  8230.& -16.\\
         &$5.0 \, \times \, 10^{-11}$&  8690.&  7460.& -15.\\
         &{\% $\Delta (\rho)$}          &     -24.&     -24.&       \\ \hline 
 B5V&$5.0 \, \times \, 10^{-12}$&  9710.&  7970.& -20.\\ 
         &$1.0 \, \times \, 10^{-11}$&  9290.&  7760.& -18.\\
         &$2.5 \, \times \, 10^{-11}$&  8450.&  7230.& -16.\\
         &$5.0 \, \times \, 10^{-11}$&  7660.&  6760.& -12.\\
         &{\% $\Delta (\rho) $}         &     -24.&     -16.&        \\ \hline 
\end{tabular}
\end{center}
\end{table}

Taking all the results into consideration, we see that while the bulk of the disk becomes cooler, 
there is evidence of heating in some parts of the disk (either from the volume-weighted average temperatures or the maximum temperatures or both).  
Heating of the upper edge of the disk is seen in the vertical profiles of the B2V and the B3V models, 
as shown is Figures~\ref{B2_zprofile} and~\ref{B3_zprofile}.  
This was also seen in Figures 8, 9 and 13 - 15 in \citet{me1}.  
When density changes are combined with rotation, 
we see that the cooling associated with increasing density occurs at lower $\rho_o$ at higher rotation rates.  
This suggests that if rotation is not taken into account, the densities of these disks could be over estimated.

\subsection{Self-Consistent Vertical Hydrostatic Equilibrium}
\label{hydro}

\begin{figure}[H]
\epsscale{0.5}
\plotone{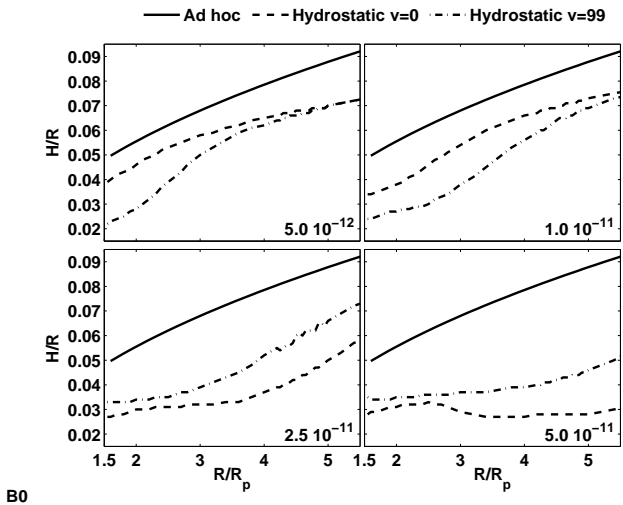}
\caption{Ratio of disk scale height to disk radius versus disk radius for the fixed density structure shown in Equation~(\ref{rho_disk}),  
and two hydrostatically converged models, one without rotation ($v_{\rm{frac}}=0$) 
and one with rotation including gravitational darkening ($v_{\rm{frac}}=0.99$) 
for the B0V model.  Four different disk densities are shown: 
$\rho_o=$ $5.0\, \times \, 10^{-12}$ (upper left); 
                      $1.0\, \times \, 10^{-11}$ (upper right); 
                      $2.5\, \times \, 10^{-11}$ (lower left); 
            and $5.0\, \times \, 10^{-11}$ g cm$^{-3}$ (lower right).  
 \label{hydrostaic}}
\end{figure}

In this section we describe the effects of rotation on disk models that have been hydrostatically 
converged, as described in \citet{hydro_paper}.  
The vertical structure of 
classical~Be~star disks are believed to be in hydrostatic equilibrium.  
Equation~(\ref{rho_disk}) governs the density structure of a hydrostatically supported 
{\emph {isothermal}} disk, but real Be star disks are not isothermal.  
Temperatures throughout the disks typically vary by factors of 2 to 3, as seen in 
Figures~\ref{B0_max_min} through~\ref{B5_max_min} and Figures~\ref{hish_5d0em12} and~\ref{hish_5d0em11}.  
This means that Equations~(\ref{rho_disk}) and~(\ref{scale_disk}) 
are inconsistent with the detailed $T(R,Z)$ distributions for the models computed from radiative equilibrium.  
Vertical pressure support in the disk requires that the vertical density profile follow $-dP/dz=\rho(z,r) g_z$, in which $g_z$ 
is the vertical component of the star's gravitational acceleration.  
The local pressure is strongly affected by the local temperature of the disk via the equation of state, $P=\rho kT/\mu $.  
This means that in order to create a density profile consistent with the temperature solution, 
an additional loop is required within the {\sc Bedisk} code to enforce hydrostatic 
equilibrium in each column, as described by \citet{hydro_paper}.  
Because of the extra loop, a converged model takes approximately ten times longer to calculate than an unconverged model.  

Figure~\ref{hydrostaic} shows the ratio of the disk scale height to the disk radius, $H(R)/R$, versus disk radius, $R$, 
in the inner disk ($r \le 5 r_{\rm {p}}$) for B0V models of both the fixed density structure described by Equation~(\ref{rho_disk}) 
and those that have been hydrostatically converged.  
$H(R)$ is the height where the density drops by a factor of $1/e$.  
For an isothermal disk, $H(R)/R$ is equal to the ratio of the sound speed, $c_{\rm{sound}}=\sqrt{P/\rho}$, 
to the Keplerian orbital speed, $v_{\rm{orbit}}=\sqrt{G M /R}$.  
Because of the nature of the fixed isothermal density structure, 
it always produces the same $H(R)/R$ profile that increases as $R^{{1}/{2}}$ regardless of $\rho_o$ or $n$; 
as it is a function of only $T_{\rm iso}$.  
In the current set of models, $H(R)/R$ is always larger for the fixed structure compared to the hydrostatically converged structure because 
we have adopted the usual result that $T_{\rm iso}=0.6 T_{\rm eff}$.  
However this temperature is often higher than the temperatures actually found in the mid-plane of dense disks.  
Therefore, Equation~(\ref{rho_disk}) over-estimates the pressure support of the mid-plane, 
(see \citet{hydro_paper} for more details).  
The difference between $T_{\rm{iso}}$ and the actual disk temperature 
becomes smaller as the disk radius increases and this is why the difference in $H(R)/R$ 
between the hydrostatically converged models 
and those with the fixed density structure decreases with increasing radius.  
The difference in $H(R)/R$ between the fixed structure and the 
hydrostatically converged models also increases as $\rho_o$ increases 
because the temperatures in the mid-plane decrease strongly as $\rho_o$ increases.  
Note that there is no single choice for $T_{\rm iso}$ that would be able to match the correct 
vertical density structures of the hydrostatically converged models shown in Figure~\ref{hydrostaic}.  

We now demonstrate that rotation, in addition to density, affects the scale heights computed for these disks.  
Figure~\ref{hydrostaic} demonstrates that  $H(R)/R$ for $v=0.99 \, \, v_{\rm{frac}}$ is always lower than that for $v=0$ for the same radius and density.  
This is because rapidly rotating models have cooler temperatures in the mid-plane, 
resulting in smaller scale heights: rapid rotation causes models with thinner disks.  
The model with $\rho_o=5.0 \, \times \, 10^{-11}$ g cm$^{-3}$ and $v=0.99\,v_{\rm{crit}}$ 
has a very thin disk with a maximum value in the inner disk  of $H(R)/R=0.033$ at 2.6 stellar radii 
and a minimum of $0.027$ at 3.75 stellar radii.  

Table~\ref{hydro_table} shows the density-weighted average temperatures without rotation and with rotation at $0.99 \, v_{\rm{frac}}$ 
for the fixed density structure and the hydrostatically converged structure 
for the four different densities shown in Figure~\ref{hydrostaic}.  
Increasing density or rotation causes a decrease in the density-weighted average temperature with or without hydrostatic equilibrium.  
Requiring self-consistent hydrostatic equilibrium reduces the effects of changing rotation and density 
and also increases the temperatures found in the dense B0V models.  

\vspace{-0.1in}
\begin{table}[H]
\caption[]{Density-weighted average temperatures (K) with and without self-consistent hydrostatic equilibrium  \label{hydro_table}}
\vspace{-0.08in}
\begin{center}
\begin{tabular}{c c c c c c c}
\hline \hline
        B0V              &\multicolumn{3}{c} {Fixed}   & \multicolumn{3}{c}      {Hydro}  \\  
$\rho_o$ (g cm$^{-3}$)& $v_{\rm{frac}}=0.$& $v_{\rm{frac}}=0.99$&\% $\Delta (v_{\rm frac})$  &$v_{\rm{frac}}=0.$& $v_{\rm{frac}}=0.99$&\% $\Delta (v_{\rm frac})$   \\     \hline
$5.0\, \times \, 10^{-12}$ & 15200& 14250&  -6&15170& 14300&   -6 \\
$1.0\, \times \, 10^{-11}$ & 15190& 13920&  -9&15120& 14070&   -7 \\
$2.5\, \times \, 10^{-11}$ & 14520& 13010&-11&14590& 13500&   -8 \\
$5.0\, \times \, 10^{-11}$ & 13280& 10780&-21&14020& 12570& -11 \\
        {\% $\Delta (\rho)$} &    -14  &     -28 &      &      -8 &      -13&        \\   \hline 
\end{tabular}
\end{center}
\end{table}

\subsection{Different forms of gravitational darkening}
\label{diff_grav_dark}

\begin{figure}[H]
\epsscale{0.5}
\vspace{-0.08in}
\plotone{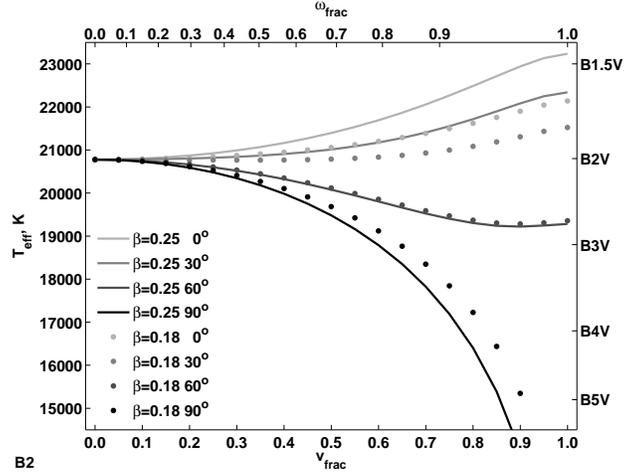}
\vspace{-0.08in}
\caption{Surface temperature versus rotation rate 
for four different co-latitudes, $0^{o}$, $30^{o}$, $60^{o}$, and $90^{o}$, for the B2V model.  
The solid lines indicate $\beta=0.25$ while dots indicate $\beta=0.18$.  
The lower horizontal axis indicates the fractional rotational velocity 
at the equator and the upper horizontal axis indicates the corresponding fractional angular 
velocity. \label{beta_0p18_theory}}
\end{figure}

As described in \citet{vanBelle2012}, gravitational darkening has been interferometrically confirmed, 
but the difference in brightness over the stellar disk is not as strong as 
that predicted by the standard formulation of gravitational darkening presented by \citet{Collins1963}.  
This formulation states that 
\vspace{-0.08in}
\begin{equation}
F=C_{\omega} g^{\, 4 \beta} \, , 
\end{equation}
\vspace{-0.08in}
where $F$ is the local radiative flux and $C_{\omega}$ is constant across a star and 
determined by the luminosity such that 
\vspace{-0.08in}
\begin{equation}
C_{\omega}= L_{\omega}/{\int g^{4 \beta} dA} \, ,  
\end{equation}
\vspace{-0.08in}
where the integration is over the surface of the star.  
The luminosity, $L_{\omega}$, can be treated as a constant or as a function of rotation.  
The canonical value of $\beta$, defined in \citet{von_Zeipel_1}, is $1/4$, but 
observations suggest a range of values between $0.25$ and $0.19$ for B~type stars \citep{vanBelle2012}.  
\citet{claret12} presents a calculation for the variation of $\beta$ with local temperature 
and optical depth  in a rotating star of 4 $M_{\odot}$ and describes departures from von Zeipel 
in the upper layers where $\beta$ can be as low as  $\sim \!  0.18$ 
when energy transport is radiative.
Figure \ref{beta_0p18_theory} shows the variation in temperature with rotation for $\beta=0.18$.   
and $0.25$ at four different stellar co-latitudes.  
The constant $C_{\omega}$ was recomputed in each case.  
Using $\beta=0.18$ weakens both the temperature increase in the polar region and 
the temperature decrease in the equatorial regions, reducing temperature difference between the equator and the pole.

\begin{figure}[H]
\epsscale{0.5}
\vspace{-0.08in}
\plotone{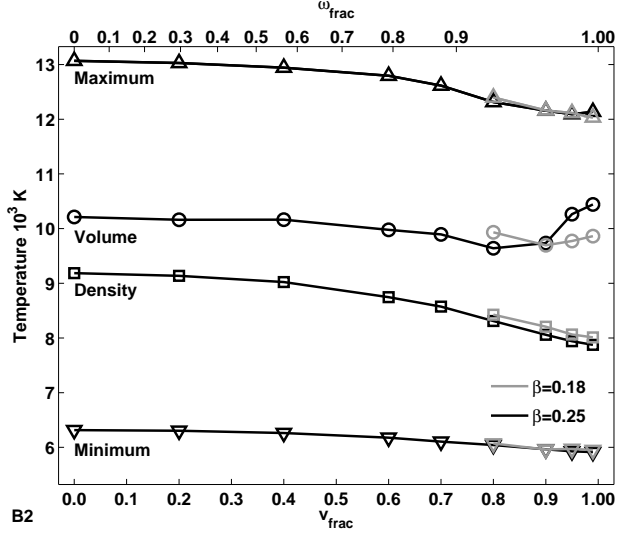}
\vspace{-0.08in}
\caption{Change in disk temperature diagnostics with increasing rotation 
for the B2V model with $\rho_o=5.0\, \times \, 10^{-11}$ g cm$^{-3}$.  
The lines marked with upward pointing triangles indicate the maximum temperature and 
the lines marked with downward pointing triangles indicate the minimum temperature.  
The lines marked with circles indicate the volume-weighted average temperature and 
the lines marked with squares indicate the density-weighted average temperatures.  
The black and grey lines show models with $\beta=0.25$ and $\beta=0.18$ respectively.  
The lower horizontal axis indicates the fractional rotational velocity at the equator 
and the upper horizontal axis indicates the corresponding fractional angular velocity.  \label{beta_0p18}}
\end{figure}

We choose $\beta=0.18$ as a lower limit and re-ran {\sc Bedisk} for the four highest rotation rates assuming the B2V star surrounded by a disk with 
$\rho_o=5.0 \times 10^{-11}$ g cm$^{-3}$.  
The results are shown in Figure~\ref{beta_0p18}.  
Lowering $\beta$ reduces the strength of gravitational darkening 
and the corresponding effect of the stellar flux on the temperatures in the disk.  
The density-weighted average temperature decreases with rotation and this decrease is less with $\beta=0.18$.  
The volume-weighted average temperatures in the disk have a more complex behaviour, 
initially decreasing then increasing with rapid rotation, as $\beta$ is reduced to $0.18$.  
The initial drop in temperature becomes smaller and the increase in temperature at higher rotation rates is also smaller.  
The value of $\beta$ does not effect the maximum or minimum temperatures found in the disk.  

The discrepancy in $\beta$ between theory and observation has motivated a closer look at gravitational darkening and possible refinements to the theory.  
\citet{Lara_2011} presents such a refinement.  
Both \citet{Lara_2011} and the canonical method begin with the expression for radiative energy transport.    
The equation of hydrostatic equilibrium is used to relate the radiative flux to the local gravity, resulting in expressions of the form, 
${F_{\rm{rad}}}\propto{{g}}$, which becomes $T_{\rm{eff}}\propto{|{g}|}^{1/4}$ using the Stephan-Boltzmann law \citep{von_Zeipel_1,Clayton}.
The difference in these approaches is how the other terms in the equation are handled.  
\citet{von_Zeipel_1} treats the equipotentials as isobaric surfaces which implies that 
all other terms in the equation are constant over the surface of a rotating star 
(for this type of argument see \S 6.8 of \citet{Clayton}).  
This leads to a contradiction: the gas temperature is taken as a constant across the surface of a rotating star, 
while the effective temperature decreases with increasing stellar co-latitude.  
Although this is not truly a paradox (as the effective temperature and the gas temperature are not the same quantities), 
it would be reasonable for a higher radiative flux at the poles to produce higher local temperatures.  
 
Alternatively, \citet{Lara_2011} make no assumptions about the other terms necessarily being constant, letting 
${F_{\rm{rad}}}= f(r,\theta){{g}}$.  The unknown function, $f(r,\theta)$, is found 
by requiring that $\nabla \cdot F_{\rm{rad}}=0$.  Using the Roche model and solving the resulting 
partial differential equation gives 
\begin{equation}
\label{ELR1}
T_{\rm{eff}} ={ (\frac{L_{\omega}}{\sigma GM})}^{1/4} \sqrt{\frac{\tan \theta_w}{\tan \theta}} {g^{1/4}} \, , 
\end{equation}
where $\theta_w$ is defined by the requirement that 
\begin{equation}
\label{ELR2}
\cos \theta_w+\ln{\tan \frac{\theta_w}{2}}=
\frac{1}{3} x^3 w^2 \cos^3 \theta+\cos \theta+\ln{\tan \frac{\theta}{2}} \, .  
\end{equation}
In Equation~\ref{ELR2}, $\theta$ is the spherical coordinate, 
$x$ is a scaled radius $r/r_{\rm eq}(\omega_{\rm star} )$, 
and $w$ is a different fractional angular velocity given by 
$\omega_{\rm star} \sqrt{r^3_{\rm eq} (\omega_{\rm star} )/G M}$ 
which is {\em not} $\omega_{\rm frac}$.  
Equations of this form which describe a variable $f(\theta)$ are sometimes written
in the equivalent form of $T=f_o g^{\beta_o +\delta(\theta) }$ ,  
where $f_o$ is now constant with all variation accounted for in $\delta( \theta)$.  
Gravity darkening laws expressed in this form are found in \citet{claret12} and \citet{zorec}.

\begin{figure}[H]
\epsscale{0.45}
\vspace{-0.1in}
\plotone{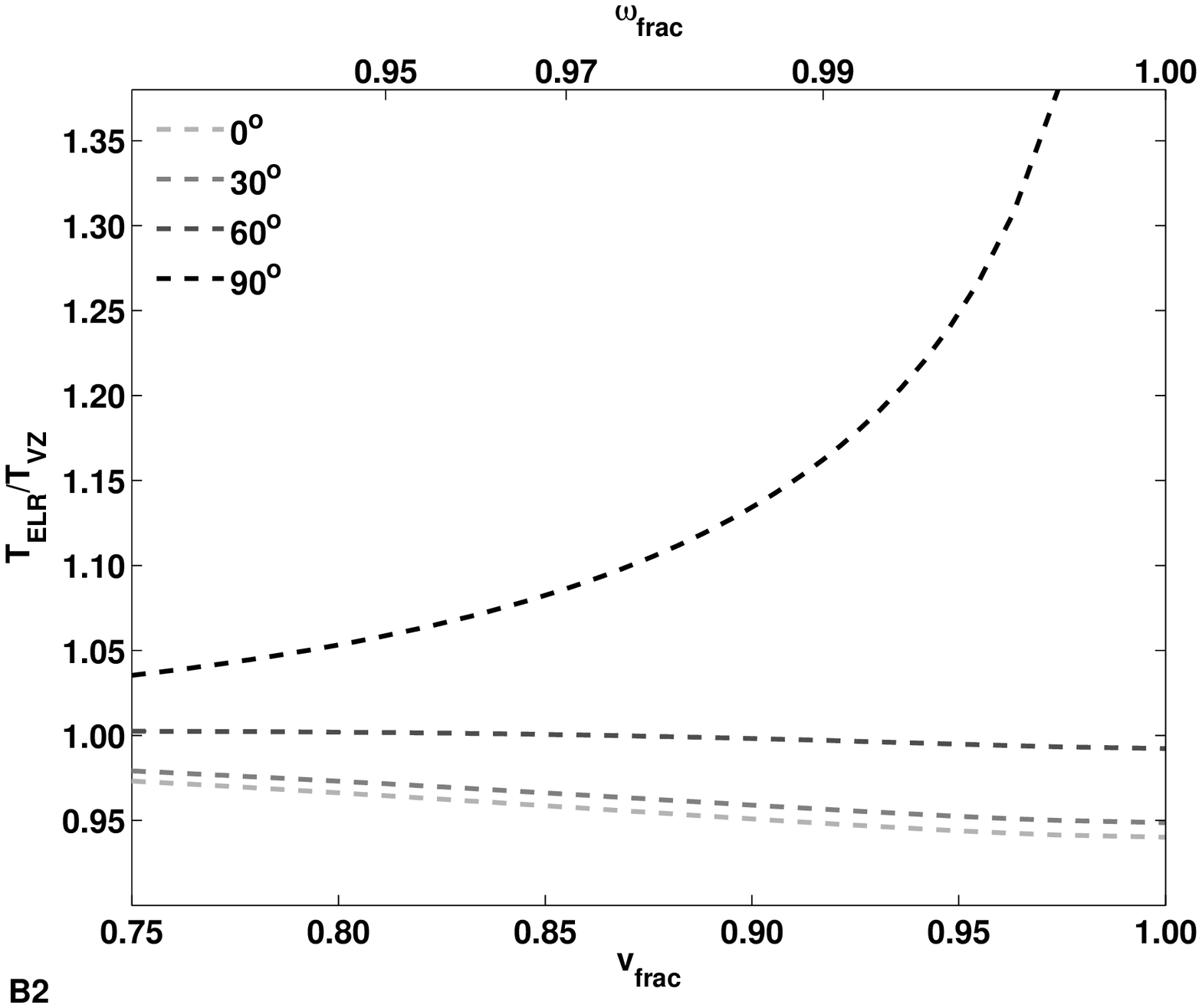}
\vspace{-0.1in}
\caption{Ratio of the temperatures predicted by \citet{Lara_2011}, $T_{\rm{ELR}}$, 
to those of standard gravitational darkening, $T_{\rm{VZ}}$, \citep{von_Zeipel_1,Collins1963}
versus rotation rate for four different co-latitudes, $0^{o}$, $30^{o}$, $60^{o}$, and $90^{o}$.  
The lower horizontal axis indicates the fractional rotational velocity 
at the equator and the upper horizontal axis indicates the corresponding fractional angular velocity.  
For readability we only show rotation rates faster than $0.75 v_{\rm{frac}}$, since this 
ratio is $\approx1$ for low rotation rates.  
 \label{lara_theroy}}
\end{figure}
\vspace{-0.1in}

\begin{figure}[H]
\epsscale{0.5}
\vspace{-0.08in}
\plotone{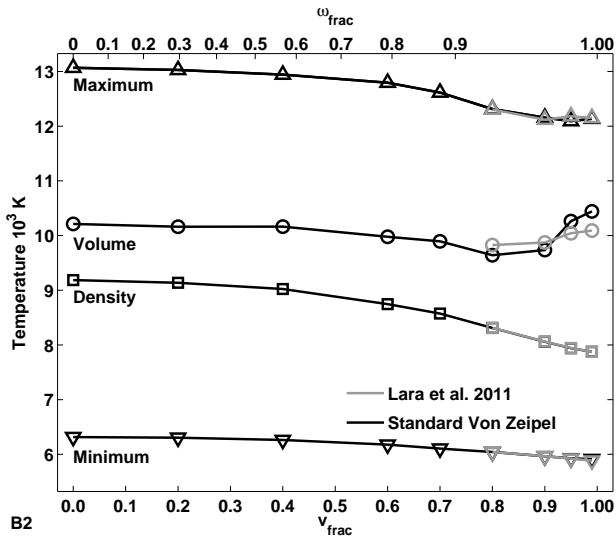}
\vspace{-0.08in}
\caption{ Same as Figure~\ref{beta_0p18} except the black lines show standard gravitational darkening with $\beta=0.25$ 
while the grey lines show results using \citet{Lara_2011}.  \label{lara_results}}
\end{figure}

Equation~(\ref{ELR1}) predicts lower polar temperatures and higher equatorial temperatures than \citet{von_Zeipel_1}.  
This is similar to, but not the same as, reducing $\beta$.  
At slow rotation rates there is very little difference between the temperatures predicted by \citet{Lara_2011} and \citet{von_Zeipel_1}, 
but the differences increase with rotation and the ratio between them actually diverges at 
the equator for critical rotation because while both functions find a temperature of zero at the equator for critical rotation, 
the prediction of \citet{Lara_2011} goes to zero with rotation more slowly (see Figure~\ref{lara_theroy}).  
Figure \ref{lara_results} shows the effect of using the Equation~(\ref{ELR1}) compared to traditional gravity darkening of \citet{von_Zeipel_1} and \citet{Collins1963}.  
It is similar to Figure~\ref{beta_0p18}, predicting a weakening of the effects of gravitational darkening.  
However, Equation~(\ref{ELR1}) reduces the heating of the pole more than the cooling of the equator, 
in comparison to simply reducing $\beta$.  
This is why using Equation~(\ref{ELR1}) does not effect the density-weighted average temperatures as much as setting $\beta=0.18$.  
Using Equation~(\ref{ELR1}) has a similar effect on the volume-weighted average temperatures as setting $\beta=0.18$, 
reducing the size of the initial drop in temperatures as well as reducing the heating at high rotation rates 
but the effect of setting $\beta$ to $0.18$ is stronger.

\section{Conclusions}
\label{conclusions}

Both density and rotation significantly affect disk temperatures.  
This is clearly seen in the density-weighted average temperatures in 
Figures~\ref{B0_max_min} through \ref{B5_max_min} and in Table~\ref{density_table}.  
Both cause disks to become cooler.  
Density is a stronger controller of the temperatures of these disks than rotation.  
This is as expected.  
Increasing the optical depth through the disk decreases the amount of photoionization radiation able to penetrate the disk by a large factor, 
especially in the mid-plane where the bulk of the material is located.  
Alternatively, rotation does not change the luminosity of the star, but only redirects it away from the mid-plane.  
The effects of rotation on the thermal structure of disks are not noticeable below $0.20 \, v_{\rm{frac}}$.  
For moderate rotation, $0.20 \, v_{\rm{crit}}$ to $0.60 \, v_{\rm{crit}}$, there are small but noticeable changes in the disk temperatures.  
From $0.60\, v_{\rm{crit}}$ to $0.80 \, v_{\rm{crit}}$ the effects of rotation become stronger as the equator cools and 
from $0.80\, v_{\rm{crit}}$ to $0.99 \, v_{\rm{crit}}$ the stellar pole becomes hot enough to influence the disk while the equatorial cooling continues.  

Increasing rotation from zero to $0.80 \, v_{\rm{frac}}$ can have the same effect on the density-weighted average temperatures 
as increasing the density by a factors of 1.5 to 5, depending on the model.  
The effects become even larger closer to critical rotation.  
Increasing rotation from zero to $0.99 \, v_{\rm{frac}}$ can have the same effect on the density-weighted average temperatures as 
increasing the density by 2.5 to 7.5 times, depending on the model.  
Therefore, while not as strong a temperature controller as density, 
rotation can be very significant for moderate to strong rotation and should not be neglected.  
Because classical Be~stars rotate faster than $0.40 \, v_{\rm{frac}}$, and many rotate faster than $0.80 \, v_{\rm{frac}}$ \citep{cra05}, 
gravitational darkening should be included in models of classical Be~stars disks .  

With increasing rotation and density, Be star disks become less isothermal.  
This is can be clearly seen in Figures~\ref{hish_5d0em12} and~\ref{hish_5d0em11}.  
These histograms show an increase in the amount of very cool gas as rotation increases, and 
at high rotation rates, there is an increase in the amount of hot gas also.  
Increasing the density of these disks increases the amount of cool gas, but does not significantly increase the temperatures in the upper disk.  
Therefore, only strong rotation causes an noticeable increase in the temperatures of the upper disk due to the hotter stellar pole.  

Classical Be~stars are known to be rapid rotators with geometrically thin disks \citep{porter}.  
The inclusion of gravitational darkening in hydrostatically converged models shows 
that these disks are predicted to be very thin around rapidly rotating stars because 
the large equatorial cool region reduces pressure support in the vertical direction.  
The scale heights predicted for the densest model considered are very small, with $H/R$ reaching as low as $0.027$ at $3.75$ stellar radii.  

Unsurprising, using weaker forms of gravitational darkening weakens the effects of gravitational darkening on the disk.  
Simply reducing $\beta$ decreases the effects of rotation on both the volume-weighted average temperatures and 
the density-weighted average temperatures in the disk, but also introduces a new free parameter.  
The formulation of \citet{Lara_2011} is a excellent alternative as it offers a physical explanation, 
avoids adding an artificial parameter, and is an algebraic solution so that 
it can be utilized for any star and any rotation rate lower than critical.  
Replacing standard gravitational darkening with the formulation of \citet{Lara_2011} does not change the 
density-weighted average temperatures in the disk very much, but it does reduce the 
heating of the upper disk due to the stellar pole.  
There are other physical phenomena which could further effect the 
temperature structure of a rotating star, such as horizontal transport of energy as 
described in \citet{had} which would re-distribute energy from the bright poles to the 
dimmer equator, but this effect is expected to be small.  

In the future, we plan to produce observables for our rotating Be~star disk models.  
It will be interesting to see how rotation and density changes effect the colours and H$\alpha$ emission from these systems, 
which are often used as the primary diagnostics of our circumstellar disk.  
For disks with large optical depths, temperature changes can strongly influence disk emission so effects due to rotation are likely to be important.  

\acknowledgements 
This research was supported in part by NSERC, 
the Natural Sciences and Engineering Research Council of Canada.  
MAM acknowledges the receipt of an Ontario Government Scholarship that funded 
part of this work.  

We would like to thank Francisco Espinosa Lara and Michel Rieutord 
for helpful comments on their formulation of gravitational darkening.  

We would also like to thank the referee, J. Zorec, for providing helpful comments.

\end{document}